\documentclass[10pt]{article}
\usepackage[utf8]{inputenc} 
\usepackage[T1]{fontenc}    

\usepackage{dsfont}
\usepackage{multirow}
\usepackage[utf8]{inputenc}
\usepackage[T1]{fontenc}
\usepackage{textcomp}
\usepackage[top=1in,bottom=1in,left=1in,right=1in]{geometry}
\usepackage{graphicx}

\newcommand{\pr}{\mathrm{Pr}}
\newcommand{\one}[1]{\mathds{1}{\left\{#1\right\}}}
\newcommand{\pmtwo}{{PM$_{2.5}$ }}
\newcommand{\pmten}{{PM$_{10}$ }}

\usepackage[dvipsnames]{xcolor} 
\usepackage{algorithm}
\usepackage{algpseudocode}
\usepackage{placeins}
\usepackage{amsmath} 
\usepackage{amssymb} 
\usepackage{bm} 
\usepackage{amsmath, amssymb, bm}
\usepackage{enumitem}
\setlist[itemize]{leftmargin=*, itemsep=0.2em, topsep=0pt, parsep=0pt}

\usepackage{booktabs} 
\usepackage{diagbox} 
\usepackage{float} 
\usepackage{siunitx}

\usepackage{listings} 
\usepackage[round, comma]{natbib}
\usepackage{makecell}
\usepackage{subfig} 
\usepackage[font=small,format=plain,justification=centering]{caption}
\newenvironment{subfigure}[2][]{\begin{minipage}[b]{#2}}{\end{minipage}}

\usepackage{tabularx} 
\usepackage{array}  
\usepackage{booktabs}  
\newcolumntype{Y}{>{\raggedright\arraybackslash}X}
\usepackage{amsmath}
\usepackage[english]{babel}

\usepackage{abstract}

\begin{document}
\pagestyle{plain}
\begin{center}

    {\Large\bfseries Joint Classification of Haze and Dust Events Using Factorial Hidden Markov Model Framework\par}
    \vspace{2em}

    {Tianhao Zhang\textsuperscript{$\dagger$}, Yixin Zhang\textsuperscript{$\dagger$}, Liang Guo\textsuperscript{$\dagger$}, Xiaoqiang Wang\textsuperscript{$\dagger$,$\ast$} \par}
    \vspace{1em}

    \small 
    \textsuperscript{1}School of Mathematics and Statistics, Shandong University, Weihai, Shandong, 264209, China\par
    \textsuperscript{*}\textit{Address for correspondence:} Xiaoqiang Wang, School of Mathematics and Statistics, Shandong University, Weihai, Shandong, 264209, China.\texttt{xiaoqiang.wang@sdu.edu.cn}
    \par
    \vspace{1.5em}

    {August 21, 2025\par}\vspace{0.5em}
\end{center}

\begin{abstract}
Haze and dust pollution events have significant adverse impacts on human health and ecosystems. Their formation-impact interactions are complex, creating substantial modeling and computational challenges for joint classification. To address the state-space explosion faced by conventional Hidden Markov Models in multivariate dynamic settings, this study develops a classification framework based on the Factorial Hidden Markov Model. The framework assumes statistical independence across multiple latent chains and applies the Walsh-Hadamard transform to reduce computational and memory costs. A Gaussian copula decouples marginal distributions from dependence to capture nonlinear correlations among meteorological and pollution indicators. Algorithmically, mutual information weights the observational variables to increase the sensitivity of Viterbi decoding to salient features, and a single global weight hyperparameter balances emission and transition contributions in the decoding objective. In an empirical application, the model attains a Micro-F1 of 0.9459; for the low-frequency classes Dust prevalence below 1\% and Haze prevalence below 10\%, the F1-scores improve from 0.19 and 0.32 under a baseline FHMM to 0.75 and 0.68. The framework provides a scalable pathway for statistical modeling of complex air-pollution events and supplies quantitative evidence for decision-making in outdoor activity management and fine-grained environmental governance.
\end{abstract}


\section{Introduction}

Global studies indicate that outdoor air pollution contributes to approximately 3.3 million premature deaths annually, with Asia—especially China—being the most severely affected region \citep{Lelieveld2015}. Among various types of air pollution events, haze and dust storms are particularly prominent and hazardous, posing serious threats to public health and ecosystems while also causing significant economic losses through disrupted production activities \citep{faridi2021economic}. Haze typically consists of a complex mixture of pollutants originating from industrial emissions, transportation, combustion processes, and mineral dust sources, predominantly composed of particulate matter such as \pmtwo and \pmten \citep{novack2020anthropogenic}. Systematic reviews show that an increase of 10~$\mu$g/m$^3$ in \pmtwo is associated with a 16\% increase in the relative risk of lung cancer, while the same increase in \pmten corresponds to a 23\% increased risk \citep{ciabattini2021systematic}. Moreover, \citet{adami2022association} demonstrated that chronic exposure to levels exceeding protective thresholds is associated with a 10\% higher risk of developing immune-mediated inflammatory diseases (IMIDs). 

Dust events, on the other hand, can cause \pmten concentrations to exceed 7400~$\mu$g/m$^3$ in a short period, as observed in Beijing during extreme sandstorm episodes, thereby triggering acute cardiopulmonary stress \citep{Li2024EPFRSDS}. According to \citet{Zhang2023SDSmortality}, the nationwide mortality risk in China increases significantly on dust storm days - by 7.49\% for ischemic stroke, 8.90\% for respiratory diseases, 12.51\% for chronic lower respiratory diseases, and 11.55\% for chronic obstructive pulmonary disease (COPD). Dust exposure also poses substantial health risks for pregnant women. A one interquartile range (3.9~$\mu$g/m$^3$) increase in \pmtwo during pregnancy is associated with a significant decrease in glomerular filtration rate (GFR), by 0.54~mL/min/1.73~m$^2$, and as much as 1.33~mL/min/1.73~m$^2$ in late pregnancy \citep{ZHAO2020105805}.

Regarding their origins, haze primarily results from the accumulation of particulate matter (PM) and gaseous precursors, along with secondary aerosol formation, whereas dust arises from soil erosion driven by strong winds. Despite their distinct underlying physical mechanisms, these two phenomena exhibit complex interactions during their transport and evolution. Observational studies suggest that during dust-haze convergence processes, dust transport can alter and disrupt the original haze aerosol trajectory and enhance vertical movement. A decreasing \(\text{PM}_{10}/\text{PM}_{2.5}\) ratio indicates a shift in the coarse-to-fine particle structure \citep{zhu2022haze2dust}. Simultaneous MAX-DOAS observations in Beijing reveal a reversal in the correlation between atmospheric water vapor (\(H_2O\)) and aerosol extinction index (AE): during haze episodes, correlation coefficients ($r$) increase while slope magnitudes ($|k|$) decrease; the opposite holds for dust episodes. Similarly, aerosol optical depth (AOD) and water vapor vertical column density (\(H_2O\) VCD) co-vary during haze events but diverge during dust events, indicating the modulating effects of dust input and atmospheric recirculation on the water vapor and aerosol fields \citep{ren2021characterization}. Furthermore, source apportionment studies show that secondary aerosols contribute 33.8\% and 30.8\% to the \(\text{PM}_{2.5}/\text{PM}_{10}\) ratio during haze-to-dust and dust swing phases, respectively, consistent with the aforementioned transport and mixing characteristics \citep{wang2022dust}.

The physical and chemical complexities noted above result in both overlapping and distinct health impacts of haze and dust events. Epidemiological evidence shows that both significantly increase the risks of respiratory and cardiovascular diseases, yet the underlying toxicological mechanisms, particle reactivity, and meteorological drivers differ. Without a clear distinction between the two, misclassification may occur, leading to flawed causal interpretations and ineffective control strategies \citep{Qiu2021,Lin2022}.

However, accurately classifying and jointly analyzing haze and dust events presents several challenges. First, their particle size distributions overlap substantially, and certain dust events are mixed with combustion or industrial emissions, making it difficult to distinguish pollution types using particle size or single chemical markers alone \citep{hinds2022aerosol,Zhang2023SDSmortality}. Second, haze and dust are triggered by complex, often interrelated meteorological drivers. For instance, high humidity enhances secondary aerosol formation and haze development \citep{kong2020water_content_haze,Zhang2015HazeVisibility}, while low soil moisture, high temperatures, and strong winds significantly lower the threshold for dust emissions \citep{Yang2019DustThreshold,qin2022study}. Additionally, dust particles can adsorb other atmospheric pollutants during transport, forming composite pollutants with combined optical and toxic properties, thereby further complicating classification \citep{Li2024EPFRSDS,hinds2022aerosol,wang2022dust}. Therefore, developing a statistical framework capable of simultaneously processing multi-dimensional observational data, dynamic meteorological conditions, and overlapping pollution sources is essential for improving classification accuracy and interpretability \citep{Qiu2021,Lin2022}.

Conventional tools such as the Air Quality Index (AQI) are inadequate for addressing this classification problem. AQI is designed to convey overall health risks to the public and lacks diagnostic capabilities regarding the underlying causes of pollution events. This limitation is particularly evident during haze-dust interaction processes. \citet{wang2022dust} noted that AQI values in the “convergence zone” between haze and dust typically fall between the high values of pure dust zones and the low values of pure haze zones, exhibiting transitional behavior. During such interactions, AQI values also exhibit rapid and pronounced fluctuations, reflecting their sensitivity to pollutant concentrations but not to pollutant origins. Thus, there is an urgent need for a novel statistical model capable of jointly processing complex observational datasets under dynamically varying meteorological and source conditions.

In the atmospheric pollution modeling domain, existing approaches can be broadly categorized into deterministic and statistical models. Deterministic models focus on the physical and chemical mechanisms governing pollutant transport and transformation. While they offer insights into underlying processes, they are often computationally intensive and cost-prohibitive. In contrast, data-driven statistical models are valued for their methodological simplicity, flexibility, and, in many scenarios, higher predictive accuracy \citep{cobourn2010enhanced}. Many studies have applied regression-based methods to associate air quality with meteorological factors and forecast pollutant concentrations. For instance, some researchers have used hybrid models combining linear regression and random forests to predict daily PM\(_{10}\) concentrations \citep{karatzas2018revisiting}. However, such models are primarily designed for numerical prediction rather than for classifying the types of evolving pollution events.

Distinct from the above, Hidden Markov Models (HMMs) offer a unique perspective by inferring latent discrete states—such as pollution event types—from observed time series. As reviewed by \citet{mor2021systematic}, HMMs and their variants have become widely adopted modeling tools in climate data analysis and beyond. Recent studies have continued using HMMs to model and forecast PM\(_{2.5}\) concentrations \citep{igbawua2024bpm25}. To the best of our knowledge, no studies have applied HMMs to the joint classification of haze and dust events. The key limitation lies in the need for standard HMMs to encode all possible combinations of events (e.g., "haze-only," "dust-only," "both," or "neither") into a single, massive state space. When there are $M$ parallel latent processes each with $K$ states, the total number of combined states becomes $K^M$, resulting in exponential computational complexity and rendering inference infeasible.

To address this bottleneck, the Factorial Hidden Markov Model (FHMM) assumes conditional independence across hidden chains, dramatically reducing model complexity \citep{ghahramani1995factorial}. FHMMs offer two structural advantages: (1) they address the issue of state space explosion, and (2) they allow interpretability by modeling each pollution process (haze or dust) via separate latent chains. Building on this, we propose a complete FHMM-based framework for jointly classifying haze and dust events. Our framework leverages the efficient inference algorithm from \citet{schweiger2019factorialhmm}, while maintaining the temporal dependence modeling strength of HMMs. This structure enables efficient, decoupled classification across high-dimensional hidden states.

Accordingly, this study adopts FHMM to decouple haze and dust processes structurally, avoid exponential state space inflation, and capture the temporal dynamics of each pollution type.

The core contributions of this work are as follows:
\begin{itemize}
    \item \textbf{A novel classification framework:} We are the first to apply Factorial Hidden Markov Models (FHMM) to the joint classification of haze and dust events. By modeling haze and dust as independent Markov chains, our framework addresses the issue of state space explosion inherent in conventional HMMs.

    \item \textbf{Tailored model design and variable selection:} To effectively distinguish between two pollution types with complex causes and features, we construct a multi-dimensional observation chain incorporating PM\(_{10}\), relative humidity, wind speed, and visibility. This design enhances the model's ability to identify the distinct "fingerprints" of each pollution event.

    \item \textbf{Systematic optimization and validation:} We introduce several key methodological enhancements: (1) a semi-supervised K-means initialization of emission matrices to ensure stable EM convergence; (2) a Gaussian copula structure to capture non-linear dependencies among observations; and (3) mutual information-based weighting of variables to emphasize critical features. These strategies significantly improve performance, particularly in detecting rare dust events.
\end{itemize}

In Section 2, We present the theoretical foundations of the FHMM framework, including model architecture, mathematical formulation, and algorithmic strategies for addressing the joint classification of haze and dust. Section 3 outlines the study region, selection of observation variables, data preprocessing, and evaluation metrics. We then demonstrate the model’s performance in classifying pollution events, followed by a detailed discussion of the results. In Section 4 we summarize the core findings, reflect on the methodological contributions, and suggest potential directions for future research. And the Supplementary materials, including additional code implementations and mathematical derivations, are provided in Appendix for reproducibility.

% ------------------------ %
\section{Model Description}
\subsection{Factorial Hidden Markov Model}

The Factorial Hidden Markov Model (FHMM) is an extension of the conventional Hidden Markov Model (HMM), in which multiple independent Markov chains describe the dynamics of the hidden state processes, and the observed variables at each time step are jointly generated by the corresponding latent states \citep{ghahramani1995factorial}.

The central assumption of this model is that the hidden chains (corresponding to Haze and Dust in our study) evolve independently. While this assumption represents a simplification from a physical perspective—there are indeed complex physical couplings between haze and dust events. However, directly modeling these couplings is often computationally prohibitive or even infeasible. The key research question addressed in this study is therefore as follows: can an FHMM—based on the key assumption of statistical independence between hidden chains—still effectively distinguish between these two complex pollution events by accurately capturing their unique multidimensional meteorological and pollutant signatures (through emission probabilities), while simultaneously leveraging time dependencies (through transition probabilities)?

This work aims to validate that assumption and to explore how optimizing the model structure (e.g., through copula-based dependence modeling and mutual information-based weighting) can maximize classification performance within this simplified framework. Such efforts also lay a methodological foundation for the future development of more physically coupled models.

We assume an FHMM with $2$ independent Markov chains representing the latent dynamics of Haze and Dust, and $E$ observational chains composed of pollution and meteorological indicators. Let each chain have length $n$, and each hidden Markov chain consist of $K$ discrete states. Specifically, let $\boldsymbol Z_t = (Z_{1,t}, Z_{2,t})^\top$ denote the joint hidden state of Haze and Dust at time $t$, and $\boldsymbol{x}_t = (x_{1,t}, \ldots, x_{E,t})^\top$ be the vector of observed variables at time $t$. The joint probability under FHMM is expressed as:

\begin{equation}
	\label{eq:FHMM}
	\pr(\boldsymbol x, \boldsymbol Z) = \pr(\boldsymbol Z_{t=1}) \prod_{t=2}^n \pr(\boldsymbol Z_{{t}}\mid \boldsymbol Z_{{t}-{1}}) \prod_{t=1}^n \pr(\boldsymbol x_{{t}} \mid \boldsymbol Z_{{t}})
\end{equation}
where:

\begin{itemize}[label=•]

\item \textit{Initial state distribution.}\;
      $\displaystyle
      \Pr(\boldsymbol Z_{1})=\prod_{j=1}^{2}\prod_{k=1}^{K}
      (\phi^{j}_{k})^{\mathds{1}_{\{Z_{j,1}=k\}}}$\\[0em]
      Here, $\phi^{j}_{k}$ denotes the initial probability that the
      $j$-th hidden chain starts in state~$k$, and
      $\mathds{1}_{\{\cdot\}}$ is the indicator function.

\item \textit{Transition probabilities.}\;
      $\displaystyle
      \Pr(\boldsymbol Z_{t}\mid\boldsymbol Z_{t-1})
      =\prod_{j=1}^{2}\prod_{k,\ell}
      (A^{j}_{k\ell})^{\mathds{1}_{\{Z_{j,t-1}=k,\;Z_{j,t}=\ell\}}}$\\[0em]
      Here, $A^{j}_{k\ell}$ is the probability of moving from state
      $k$ to $\ell$ in the $j$-th hidden chain.\\[0.2em]

\item \textit{Emission distribution.}\;
      $\displaystyle
      b_{(k_1,k_2)}(\boldsymbol x_t)=
      \Pr(\boldsymbol x_t\mid Z_{1,t}=k_1,\;Z_{2,t}=k_2)$,
      \;$k_1,k_2=1,\dots,K$\\[0.3em]
      where $b_{(k_1,k_2)}(\boldsymbol x_t)$ is the joint emission
      density conditioned on hidden states $(k_1,k_2)$ at time~$t$.\\[0.15em]

\end{itemize}

For the observational variables $\boldsymbol{x} = (\boldsymbol{x}_t)_{t=1,\cdots,n}$, we tested different distributional assumptions and ultimately selected the log-normal distribution to model the four observational variables: PM$_{10}$ concentration, wind speed, visibility, and relative humidity. (The rationale for variable selection and distributional assumptions is detailed in the next section.)

However, due to limitations in observational instrumentation, the visibility data collected from meteorological stations (e.g., Beijing Capital Airport) exhibits right-censoring: all visibility values greater than or equal to 10 km are recorded as exactly 10 km, with this value being disproportionately frequent. This observed pattern clearly deviates from a standard log-normal distribution.

To address this, we decompose the observation vector into two parts: $\boldsymbol{x}_t = (\boldsymbol{x}_t^{oth}, x_t^{vis})$, where $x_t^{vis}$ is the visibility measurement and $\boldsymbol{x}_t^{oth}$ includes all other variables. Using $Z_{1,t} = \one{\text{Haze at time $t$}}$ and $Z_{2,t} = \one{\text{Dust at time $t$}}$, we construct the following ``inflated-censored log-normal mixture model'' (hereafter referred to as the inflated mixture model) for the emission distribution:

When $k_1 + k_2 \neq 0$:
\[
(\boldsymbol{x}_t \mid Z_{1,t} = k_1, Z_{2,t} = k_2) \sim \mathcal{LN}(\boldsymbol{\mu}_{(k_1,k_2)}, \Sigma_{(k_1,k_2)})
\]
That is,
\[
\log b_{(k_1,k_2)}(\boldsymbol x_t) = (2\pi)^{-\frac E2} |\Sigma_{(k_1,k_2)}|^{-1}\exp\left\{-\frac12 (\boldsymbol x_t - \boldsymbol{\mu}_{(k_1,k_2)})^\top\Sigma_{(k_1,k_2)}^{-1} (\boldsymbol x_t - \boldsymbol{\mu}_{(k_1,k_2)})\right\}
\]

When $k_1 = k_2 = 0$:
\begin{align*}
    (\boldsymbol{x}_t^{oth} \mid Z_{1,t} = 0, Z_{2,t} = 0) &\sim  \mathcal{LN}(\boldsymbol{\mu}_{(0,0)}, \Sigma_{(0,0)}) \\
    (x_t^{vis} \mid Z_{1,t} = 0, Z_{2,t} = 0) &\sim \pi_0 \one{x_t^{vis} = 10} + (1-\pi_0) \mathcal{LN}(\theta,\eta^2)
\end{align*}
Here, $0 < \pi_0 < 1$ is the mass at the inflated value, and we assume conditional independence between $\boldsymbol{x}_t^{oth}$ and $x_t^{vis}$.

In summary, we denote the complete model as $\mathrm{FHMM}(\boldsymbol{\phi}, \boldsymbol{A}, \boldsymbol{B}, \boldsymbol{M})$, where $\boldsymbol{M} = (\boldsymbol{\mu}, \Sigma, \pi_0, \theta, \eta^2)$ represents the collection of parameters for the emission distributions.

\subsection{Dependency Structure Modeling: Joint Gaussian vs. Log-Normal Copula Model Approach}
\label{sec:dependency_modeling}

After determining the marginal distributions of the observed variables, the next step is to characterize the dependency structure among them. In this study, we consider and implement two distinct modeling strategies: (1) the Joint Gaussian Model (JG), based on a covariance matrix and introduced earlier; and (2) the Log-Normal Copula Model (LNC), grounded in copula theory. The key distinction between the two approaches lies in how they address the ``inflated point'' phenomenon in the visibility data under ``clear'' weather conditions ($k_1 = k_2 = 0$). This issue arises from the empirical concentration of visibility observations at 10 km, resulting in a mixed distribution comprising a continuous component and a point mass.

For the JG model, the entire dependency structure is defined by a standard covariance matrix, which cannot, in principle, account for the interference introduced by this type of mixed distribution. To ensure model tractability, the JG model adopts a simplification that contradicts physical reality: it assumes independence between visibility and other meteorological variables under the ``clear'' state. While this assumption enables computation, it sacrifices the physical interpretability of the underlying relationships—a theoretical limitation worth scrutinizing.

To overcome this inherent limitation of the JG model, we introduce the LNC framework. The fundamental advantage of copula theory lies in its ability to factorize the joint distribution into two components: (i) the marginal distributions of individual variables, and (ii) a copula function that captures their pure dependency structure. Once this decomposition is adopted, careful consideration must be given to the choice of the copula function.

We observe that different pollution events exhibit complex and often opposing tail dependence patterns: for example, haze events tend to co-occur with low wind speeds (lower tail) and high humidity (upper tail), whereas dust events are associated with high wind speeds (upper tail) and low humidity (lower tail). This asymmetry renders any single-direction, standard asymmetric copula unsuitable across all hidden states. For this reason, we adopt the Gaussian copula as the core of the LNC model.

This choice is supported by mathematical consistency with the assumptions underlying our model. We provide a rigorous proof in the appendix that a Gaussian copula combined with log-normal marginals is mathematically equivalent to a multivariate log-normal distribution. This theoretical result substantiates our model selection and positions the LNC model as a conceptual extension of the JG model. Its key advantage lies in preserving log-normal dependency structure while decoupling marginals from dependency, thus allowing it to handle mixed distributions that the JG model cannot accommodate.

\begin{figure}[H]
    \centering
    \includegraphics[width=\textwidth]{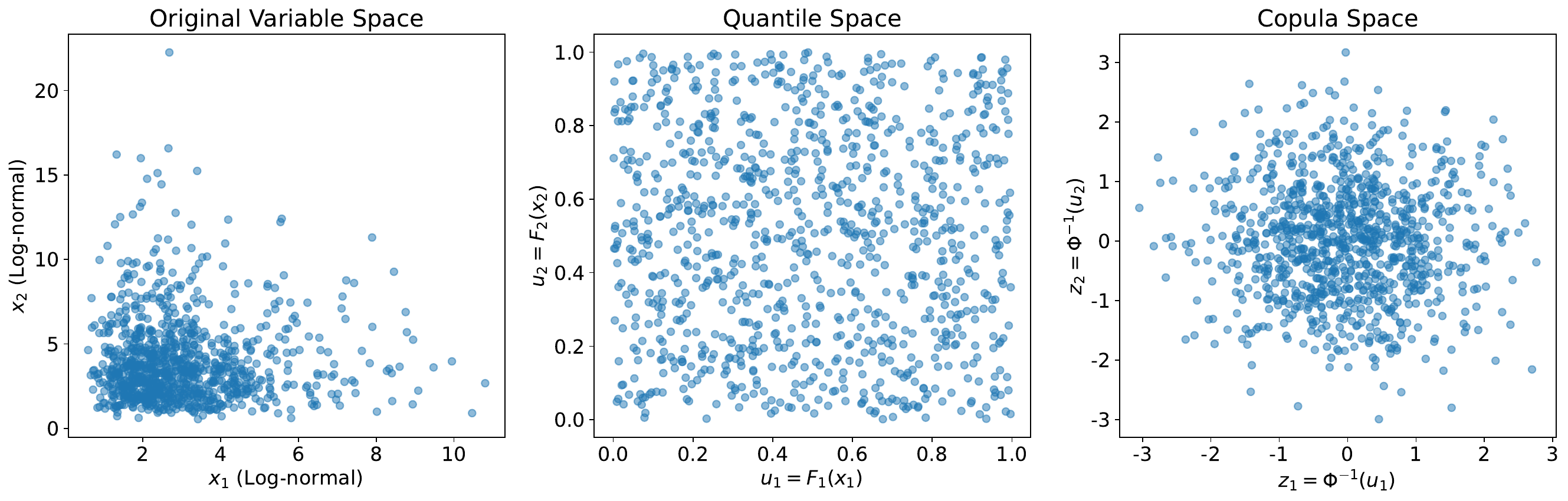}
    \caption{Illustration of the variable transformation process in the Gaussian Copula framework.(a) original variable space; (b) quantile space after probability integral transform; (c) standard normal space after inverse standard normal CDF.}
    \label{fig:Gaussian_Copula}
\end{figure}

The implementation of the Gaussian copula follows a three-step variable transformation process, illustrated in Figure~\ref{fig:Gaussian_Copula}.

First, for each variable, we apply its cumulative distribution function (CDF) $F_{x_i}$ to perform a probability integral transform: $u_{i,t} = F_{x_i}(x_{i,t})$. This maps the original variable space (left panel) into the uniform $[0,1]$ copula space (middle panel), where only the pure dependency structure remains. Then, the inverse CDF of the standard normal distribution ($\Phi^{-1}$) is applied to map the uniform variables $u_i$ to the standard normal space (right panel). In this final space, the complex nonlinear dependencies among variables are fully captured by a linear correlation matrix $\bm{R}$.

It is important to note that the purpose of these transformations is not to map the data back to the original space, but rather to compute the density of the Gaussian copula in the standard normal space. This density is then combined with the original marginal distributions (e.g., log-normal or inflated mixture) to construct the full joint distribution in the LNC model.

The resulting joint density (i.e., the corrected emission probability) for the LNC model is given by:
\begin{equation}\label{eq:LNC_density_state}
\widetilde b_{(k_1,k_2)}(\bm{x}_t)=
\lvert\bm{R}_{(k_1,k_2)}\rvert^{-\frac12}
\exp\!\left\{-\frac12\bm{Z}_t^{\top}
\left(\bm{R}_{(k_1,k_2)}^{-1}-\bm{I}\right)
\bm{Z}_t\right\}
\prod_{i=1}^{E}\!
\mathcal{LN}\!\left(x_{i,t};
\mu_{i,(k_1,k_2)},\sigma^{2}_{i,(k_1,k_2)}\right)
\end{equation}
where $\bm{Z}_t$ is the transformed vector obtained via the probability integral transform followed by the inverse standard normal CDF, $\Phi^{-1}$. To balance model flexibility with robustness in parameter estimation, we assume that all hidden states share a common global correlation matrix $\bm{R}_{\mathrm{global}}$.

In summary, we implement both the JG and LNC models in parallel. The JG model is characterized by simplicity but makes independence assumptions in specific states. The LNC model, on the other hand, corrects this limitation using copula theory and provides a more physically plausible dependency structure. A comprehensive empirical comparison of their performance will follow in later sections.

\subsection{Parameter Estimation}

Given that the dataset is annotated with explicit labels for pollution events—indicating the presence of “Haze” and “Dust” at each time step—supervised estimation can be directly employed to initialize model parameters. Specifically, the transition probability matrices are obtained by computing the empirical frequencies of state transitions from the time-ordered labeled sequence. For the emission distributions, the dataset is partitioned according to label combinations, and parameters for each subset are estimated via maximum likelihood.

For datasets without event labels, after initialization (details provided in the subsequent section), we adopt the Expectation-Maximization (EM) algorithm to obtain maximum-likelihood parameter estimates. In the Expectation (E) step, the forward-backward variables are computed as:
\begin{equation}
\begin{aligned}
\alpha_1(k_1,k_2) &\propto \phi^{1}_{k_1}\,\phi^{2}_{k_2}\, b_{(k_1,k_2)}(\boldsymbol{x}_1),\\
\alpha_t(k_1,k_2) &= \Big[\sum_{\ell_1,\ell_2}\alpha_{t-1}(\ell_1,\ell_2)\,A^{1}_{\ell_1k_1}A^{2}_{\ell_2k_2}\Big]\, b_{(k_1,k_2)}(\boldsymbol{x}_t),\\
\beta_T(\ell_1,\ell_2) &= 1,\\
\beta_t(\ell_1,\ell_2) &= \sum_{k_1,k_2} A^{1}_{\ell_1 k_1}A^{2}_{\ell_2 k_2}\, b_{(k_1,k_2)}(\boldsymbol{x}_{t+1})\, \beta_{t+1}(k_1,k_2).
\end{aligned}
\label{eq:fwd}
\end{equation}
where $\bm{x}_t$ denotes the observation vector at time $t$.

The smoothed posterior probabilities are then obtained as:
\begin{equation}
	\label{eq:postPr}
	\begin{aligned}
		\xi_t(k_1,k_2, \ell_1, \ell_2) &= 
		\frac{\alpha_t(k_1,k_2)A^1_{k_1\ell_1}A^2_{k_2\ell_2}
			b_{(\ell_1,\ell_2)}(\bm{x}_{t+1})
			\beta_{t+1}(\ell_1,\ell_2)}
		{\sum_{k_1,k_2}\alpha_t(k_1,k_2)\,\beta_t(k_1,k_2)},\\ 
		\gamma_t(k_1,k_2) &= \sum_{\ell_1,\ell_2} \xi_t(k_1,k_2, \ell_1, \ell_2).
	\end{aligned}
\end{equation}

In FHMM, the computational cost of the E-step is dominated by posterior state computations. Since the hidden chains evolve independently, the joint state $(k_1,k_2)$ can be regarded as a product state, and the overall transition matrix factorizes as the Kronecker product:
\begin{equation}
	\bm{A} = \bm{A}^1 \otimes \bm{A}^2,
	\label{eq:KrA}
\end{equation}
where $\bm{A}^1$ and $\bm{A}^2$ are the sub-chain transition matrices.

We implement the recursive matrix-vector multiplication algorithm of \citet{schweiger2019factorialhmm}, which reduces the naive complexity $\mathcal{O}(T K^{2M}) = \mathcal{O}(T K^{4})$ to:
\[
  \mathcal{O}(M T K^{M+1}) = \mathcal{O}(2 T K^{3}),
\]
with $K$ states per chain and sequence length $T$. Memory requirements are likewise reduced from $\mathcal{O}(K^{2M})$ to $\mathcal{O}(M K^{2})$, as only the sub-chain matrices need to be stored. Pseudocode for the efficient forward-backward recursion is given in Appendix~A.

In the Maximization (M) step, the parameter set $\bm{\Theta} = (\bm{\pi}, \bm{A}, \bm{B}, \bm{M})$ is updated by solving:
\[
\bm{\Theta}^{(h+1)} = \arg\max_{\bm{\Theta}}\;
\mathbb{E}\!\left[\log \Pr(\bm{x}, \bm{Z}; \bm{\Theta}) \,\middle|\, \bm{x}, \bm{\Theta}^{(h)}\right].
\]

For the Markov-chain parameters:
\[
\bm{\pi}^{(h+1)} = [\gamma_{1}(k_1,k_2)]_{K\times K}, \quad 
\bm{A}^{(h+1)} = \left[\frac{\sum_{t\geq 2} \xi_t(k_1,k_2, \ell_1, \ell_2)}{\sum_{t\geq 2} \gamma_{t}(k_1,k_2)}\right]_{K^2\times K^2}.
\]

For the emission distribution parameters:

If $k_1 + k_2 \neq 0$:
\begin{equation}
    \label{eq:muSigma}  
    \bm{\mu}_{(k_1,k_2)}^{(h+1)} = \frac{\sum_t \gamma_{t}(k_1,k_2) \bm{x}_t^{oth}}{\sum_t \gamma_{t}(k_1,k_2)}, \quad \Sigma_{(k_1,k_2)}^{(h+1)} = \frac{\sum_t \gamma_{t}(k_1,k_2) \left(\bm{x}_t^{oth} - \bm{\mu}_{(k_1,k_2)}^{(h+1)}\right)\left(\bm{x}_t^{oth} - \bm{\mu}_{(k_1,k_2)}^{(h+1)}\right)^\top}{\sum_t \gamma_{t}(k_1,k_2)}.
\end{equation}

If $k_1 = k_2 = 0$, the mean and covariance $\bm{\mu}_{(0,0)}^{(h+1)}$ and $\Sigma_{(0,0)}^{(h+1)}$ are computed analogously to \eqref{eq:muSigma} for $\bm{x}_t^{vis}$. The inflated-mixture parameters $(\pi_0,\theta,\eta^2)$ are obtained via numerical maximization:
\[
(\theta^{(h+1)}, \eta^{2,(h+1)}) = \arg\max_{\theta, \eta^2} \sum_t \gamma_t(0,0) \log\left(\pi_0 \,\mathbb{I}\{x_t^{vis} = c\} + (1-\pi_0) \,\mathcal{LN}(\theta,\eta^2)\right),
\]
where no closed form exists.

For the LNC model, the log-normal covariance $\Sigma$ is replaced by a correlation matrix $\bm{R}$. Let $\bm{Z}_{t}$ be the copula-transformed vector (Section~\ref{sec:dependency_modeling}). The global correlation matrix is estimated as:
\[
  \bm{S}_{\mathrm{global}}^{(h)}
  = \frac{\sum_{k_{1}=1}^{K}\sum_{k_{2}=1}^{K}\sum_{t=1}^{n}
      \gamma_{t}(k_{1},k_{2})\,\bm{Z}_{t}\,\bm{Z}_{t}^{\top}}
       {\sum_{k_{1}=1}^{K}\sum_{k_{2}=1}^{K}\sum_{t=1}^{n}\gamma_{t}(k_{1},k_{2})},
\]
\[
  \bm{D}_{\mathrm{global}}^{(h)} = \operatorname{diag}\!\left(\bm{S}_{\mathrm{global}}^{(h)}\right), \quad
  \bm{R}_{\mathrm{global}}^{(h+1)} = 
  \left(\bm{D}_{\mathrm{global}}^{(h)}\right)^{-\frac12}
  \bm{S}_{\mathrm{global}}^{(h)}
  \left(\bm{D}_{\mathrm{global}}^{(h)}\right)^{-\frac12}.
\]
If $\lambda_{\min}(\bm{R}_{\mathrm{global}}^{(h+1)}) < \varepsilon$, we adjust:
\[
  \bm{R}_{\mathrm{global}}^{(h+1)} \leftarrow 
  \bm{R}_{\mathrm{global}}^{(h+1)} + 
  \left(\varepsilon - \lambda_{\min}\right)_{+} \bm{I},
\]
ensuring positive definiteness. This single $\bm{R}_{\mathrm{global}}$ is then applied uniformly to all hidden-state pairs $(k_1, k_2)$.

\subsection{Mutual Information and the Viterbi Algorithm}

It is noteworthy that wind speed is one of the decisive factors influencing the formation of Dust events—without wind, dust cannot be mobilized. However, its impact on Haze events is relatively limited, primarily acting as a suppressive factor when wind speeds are excessively strong. In contrast, visibility exhibits more abrupt and pronounced fluctuations during Haze conditions compared to Dust events. This is because fine particulate matter, when combined with atmospheric moisture, significantly enhances light scattering and can sharply reduce urban visibility within a short time. Dust, by contrast, is often obstructed by urban structures and vegetation, making it less effective in rapidly lowering visibility. Given these contrasting roles of environmental variables in characterizing Haze and Dust dynamics, we introduce a structural enhancement to the traditional FHMM framework based on mutual information from information theory, and propose a refined Viterbi algorithm to improve the model’s dynamic recognition capability under complex meteorological conditions.

Mutual information quantifies the statistical dependence between continuous observations $\boldsymbol x$ and discrete hidden states $\boldsymbol Z$ using the following expression:
\[
I(\boldsymbol x; \boldsymbol Z) = \iint \pr(\boldsymbol x, \boldsymbol Z) \log \frac{\pr(\boldsymbol x, \boldsymbol Z)}{\pr(\boldsymbol x) \pr(\boldsymbol Z)} \, \mathrm d\boldsymbol x \, \mathrm d \boldsymbol Z
\]

In this study, to estimate mutual information from a given dataset $\{(\boldsymbol x_t, \boldsymbol Z_t), t=1,\ldots,n\}$, we adopt a non-parametric method based on $k$-nearest neighbors (k-NN). This approach is inspired by the work of \citet{Kraskov2004} and adapted by \citet{Ross2014} to estimate mutual information between continuous variables and discrete states. The estimation formula is given by:

\begin{equation}
\label{eq:mi_ross}
I(\boldsymbol x; \boldsymbol Z) = \psi(n) + \psi(k) - \frac{1}{n} \sum_{c' \in \mathcal{C}} N_{c'} \psi(N_{c'}) - \frac{1}{n} \sum_{t=1}^n \psi(\max(1, M_t))
\end{equation}
where:
\begin{itemize}
    \item $\psi(\cdot)$ is the digamma function, $n$ is the total number of samples, and $k$ is the predefined number of nearest neighbors (e.g., $k=3$ in this study).
    \item $\mathcal{Z}$ denotes the set of all possible discrete hidden states.
    \item $N_{\boldsymbol Z'}$ is the number of samples with hidden state $\boldsymbol Z'$.
    \item For each sample $(\boldsymbol x_t, \boldsymbol Z_t)$:
    \begin{itemize}
        \item Find the $k$-th nearest neighbor of $\boldsymbol x_t$ across the entire observation space (regardless of class), and denote the corresponding distance as $\rho_k(\boldsymbol x_t)$.
        \item Define $M_t$ as the number of other samples $\boldsymbol x_s$ (with $s \ne t$ and $Z_s = Z_t$) whose distance to $\boldsymbol x_t$ is less than $\rho_k(\boldsymbol x_t)$. We use $\max(1, M_t)$ to ensure numerical stability in the digamma function.
    \end{itemize}
\end{itemize}

This k-NN-based estimator adaptively accounts for local data density, avoiding bin width selection issues common in histogram-based methods.

Next, we define a weight matrix $W = [w_{i,(k_1,k_2)}]_{E\times K^2}$ as:

\begin{equation}
\label{eq:weight}
w_{i,(k_1,k_2)} = \Omega \cdot \frac{I(x_i; \boldsymbol{Z} = (k_1,k_2))}{\sum_{n=1}^{E} I(x_n; \boldsymbol{Z} = (k_1,k_2))}
\end{equation}
where $\Omega$ is a scaling parameter and is set to $\Omega = 1$ by default unless otherwise specified.

We then adjust the FHMM emission distribution as follows:

\begin{equation}
\label{eq:weight_emission}
b_{(k_1,k_2)}^{\langle w\rangle}(\boldsymbol x_t) = \pr^{\text{Weighted}}(\boldsymbol x_t \mid \boldsymbol Z_t = \boldsymbol Z) = \frac{\prod_{i=1}^E \left[\pr(x_{i,t} \mid Z_{1,t} = k_1, Z_{2,t} = k_2)\right]^{w_{i,(k_1,k_2)}}}{C^{\text{Weighted}}_{(k_1,k_2)}}
\end{equation}
where the normalization constant is given by:

\begin{equation}
\label{eq:emisAdj}
C^{\text{Weighted}}_{(k_1,k_2)} = \idotsint_{\mathbb{R}^{E}} \prod_{i=1}^E \left[\pr(x_{i,t} \mid Z_{1,t} = k_1, Z_{2,t} = k_2)\right]^{w_{i,(k_1,k_2)}}
\end{equation}

Note that under the inflated distribution setting, \eqref{eq:emisAdj} has no closed-form solution. We approximate this normalization term, and under our dataset, the relative error of this approximation remains below \(4 \times 10^{-4}\), which is negligible for practical purposes. Detailed derivation and error analysis are provided in the appendix.

\paragraph{On the Use of Covariance/Correlation Matrices in the Viterbi Phase}\quad

It is important to emphasize that the two emission models implemented in this study handle dependency structures differently during Viterbi decoding:

\begin{itemize}
  \item \textbf{Joint-Gaussian Model (JG):} \\
  The basic JG model cannot handle inflated mixtures. Therefore, during Viterbi recursion, only the product of marginal log-normal densities is used:
  \[
  \prod_{i} f_{LN}\left(x_{i,t}; \mu_i, \sigma_i^2 \right)
  \]
  The covariance matrix $\Sigma_{(k_1,k_2)}$ estimated during the EM step does not appear in the decoding emission score. Hence, while linear correlations between variables are considered during training (excluding the inflated visibility state), the decoding assumes conditional independence.

  \item \textbf{Log-Normal Copula Model (LNC):} \\
  In the LNC model, the Viterbi step fully utilizes the correlation matrix $\boldsymbol R_{\text{global}}$ estimated during EM. The emission log-likelihood term includes:
  \[
  -\frac{1}{2}\log|\boldsymbol R| - \frac{1}{2}\boldsymbol Z_t^\top \left(\boldsymbol R^{-1} - \boldsymbol I\right)\boldsymbol Z_t,
  \quad
  \boldsymbol Z_t = \Phi^{-1}(F_{LN}(x_{i,t}))
  \]
  This ensures consistency between training and decoding, enabling the model to fully exploit the multidimensional dependencies captured by the copula structure.
\end{itemize}

\noindent
In summary, only the LNC model maintains consistent correlation structures throughout both training and inference. The JG model, by contrast, influences parameter updates in the EM stage but degenerates to independent marginal estimation in the Viterbi decoding. This key difference will be further examined in the empirical comparison presented in the following sections.

\subsection{Pollution Event Forecasting}

Once the parameters are estimated, hidden states can be inferred from the observations and future states can be forecast. We adopt the state transition based predictor of \citet{rabiner1986introduction}:

Given observations $\boldsymbol x_1,\dots,\boldsymbol x_T$, let $\boldsymbol{\alpha}_T$ be the filtered state weights and define $\boldsymbol{\phi}_T=\boldsymbol{\alpha}_T/L_T$ with $L_T=\sum_{k}\alpha_T(k)$. With transition matrix $\boldsymbol A$ and horizon $h$, the $h$-step ahead distribution is
\[
\begin{aligned}
\mathrm{Pr}\big(\boldsymbol Z_{T+h}=\ell \mid \boldsymbol x_1,\dots,\boldsymbol x_T\big)
&= \boldsymbol{\phi}_N\,\boldsymbol A^{h}\,\boldsymbol e_{\ell}^{\top},
\end{aligned}
\]
where $\boldsymbol e_{\ell}$ is the unit vector. This yields forecasts for $\boldsymbol Z_{N+h}$ and supports posterior predictive checks by sampling $\boldsymbol Z_{N+1},\dots,\boldsymbol Z_{N+h}$ and the corresponding emissions for comparison with held out measurements.
% ------------------------ %
\section{Results and Analysis}
\subsection{Study Area}

This study focuses on Beijing, one of the most severely air-polluted cities in China and the political, cultural, and economic center of the country. Situated at the core of the Beijing-Tianjin-Hebei region, the city has a permanent population exceeding 21 million. High concentrations of industrial activity, extensive vehicle fleets, and elevated emission intensities rank among the highest nationwide. The northwest is flanked by the Taihang and Yanshan mountain ranges, whose complex topography creates unfavorable meteorological conditions—particularly reduced vertical atmospheric dispersion—that facilitate pollutant accumulation and secondary aerosol formation \citep{Ma2023NightPMES, CHEN2018290, Zhang2015HazeVisibility}.

\begin{figure}[H] 	
    \centering 	
    \includegraphics[width=0.45\textwidth]{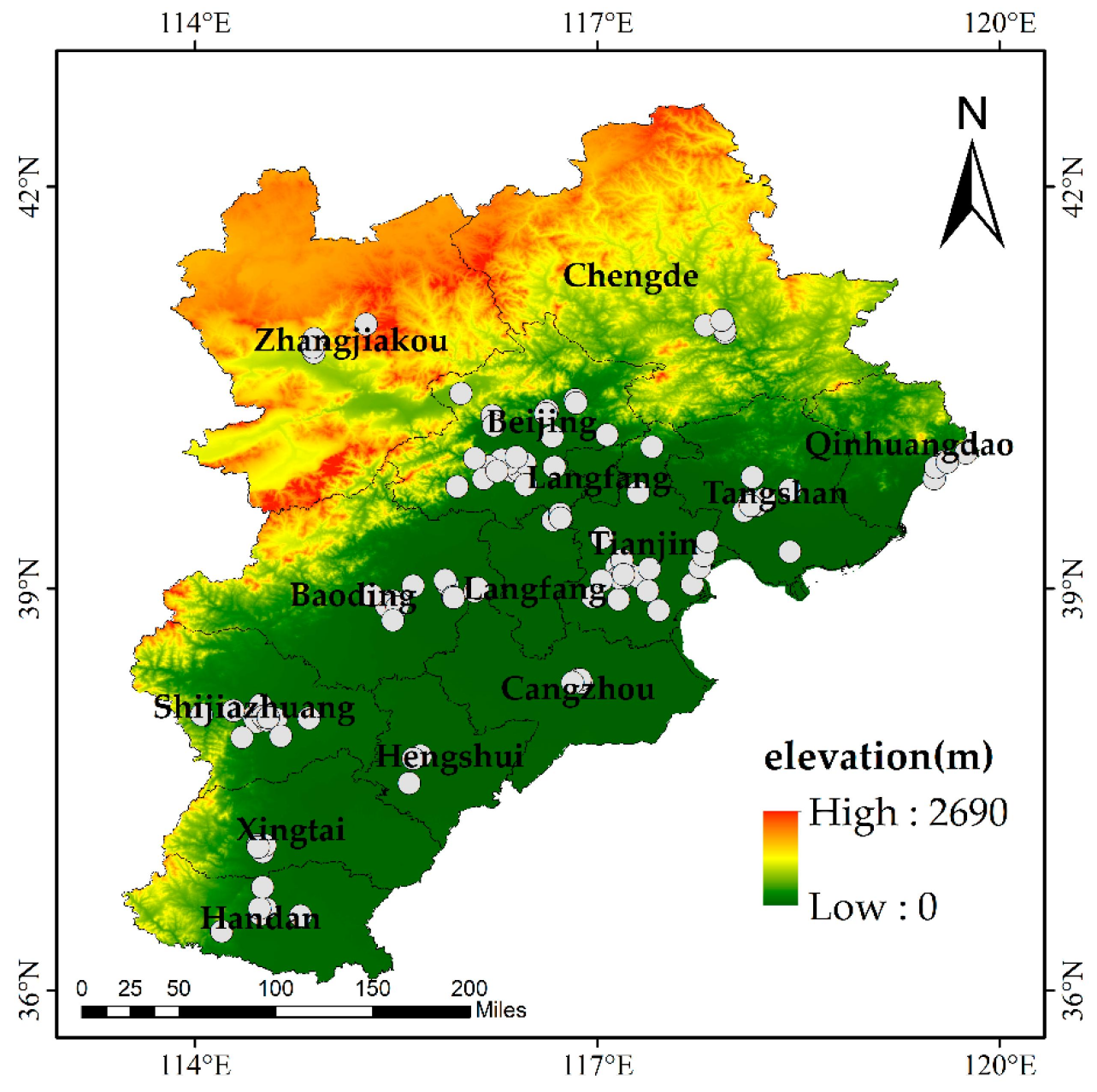} 	
    \caption{Topographic map of the Beijing-Tianjin-Hebei region. Source: \citep{Ma2023NightPMES}.} 	\label{fig:topography_map} 
\end{figure}

Beijing experiences frequent high-intensity haze and dust events, especially during winter and spring. Between 2015 and 2022, more than 20 sand and dust storm events were recorded \citep{Li2024EPFRSDS}, typically marked by abrupt increases in particulate concentrations, sharp drops in visibility, and enhanced photochemical pollution \citep{Zhang2015HazeVisibility, Jiang2024}. In extreme cases, PM$_{10}$ concentrations can exceed \(7400~\mu\text{g}/\text{m}^3\) \citep{Li2024EPFRSDS}. Epidemiological studies indicate that each \(10~\mu\text{g}/\text{m}^3\) increase in PM$_{10}$ is associated with a 0.37\% increase in emergency visits for arrhythmia and a 0.20\% increase for cerebrovascular diseases \citep{Shi2021PM10CSD}. Under dust conditions, particle oxidative potential and free radical levels rise markedly, potentially exacerbating respiratory damage \citep{Li2024EPFRSDS}. Moreover, sandstorms have been shown to adversely affect mental health and disrupt daily life, particularly during severe episodes \citep{Chang2024}.

Lidar observations in spring 2024 revealed frequent accumulation of haze aerosols—comprising desert dust, mining dust, and anthropogenic pollutants—in Beijing’s lower atmosphere. These aerosols exhibit strong hygroscopicity, complex vertical structures, fine particle sizes, and high oxidative activity, which collectively enhance deposition efficiency and biological toxicity in the respiratory tract, thereby posing compounded health risks \citep{Jiang2024}. Health-related economic losses from PM pollution in Beijing were estimated at approximately 67.9 billion RMB in 2016, with impacts concentrated in densely populated districts \citep{Chen2018b}.

Given its diverse pollution sources, enclosed terrain, dense population, and compounded health risks, Beijing offers a representative and strategically important setting for joint classification and mechanistic modeling of haze and dust pollution, providing both theoretical significance and practical relevance.

\subsection{Dataset and Model Initialization}

\subsubsection{Selection of Pollution Indicators and Dataset}

To enable robust classification of two dynamically evolving atmospheric pollution states—Haze and Dust—this study selects key air quality and meteorological indicators grounded in their distinct yet complementary physical mechanisms. These variables form a multi-dimensional observation chain within the FHMM framework.

Visibility (km), the most direct macroscopic indicator, typically declines during both Haze and Dust episodes \citep{Kong2019HazeHumidity,Zhang2015HazeVisibility} and thus serves as a primary criterion for identifying pollution events. To further distinguish between events of differing origins, two meteorological variables with contrasting indicative roles are included: wind speed (m/s) and relative humidity (RH, \%). Wind speed exhibits a threshold effect on Dust generation, requiring values above the critical entrainment velocity \citep{Yang2019DustThreshold}, while also suppressing Haze formation by enhancing near-surface ventilation \citep{Zhang2015HazeVisibility}. In contrast, RH facilitates Haze development through hygroscopic particle growth \citep{Kong2019HazeHumidity,ren2021characterization} but is generally low during Dust episodes \citep{Yang2019DustThreshold,ren2021characterization}.

PM$_{10}$ concentration ($\mu$g/m$^3$) is incorporated to represent coarse particulate loading, capturing differences in particle size distributions between event types. In dust source regions, coarse particles dominate and PM$_{10}$ levels far exceed PM$_{2.5}$ \citep{wang2022dust}, whereas in haze-affected areas, PM$_{2.5}$ is predominant but PM$_{10}$ also remains elevated. Thus, PM$_{10}$ serves both as a tracer for Dust and as a discriminator between Haze and Clear conditions.

\begin{figure}[H]
    \centering
    \includegraphics[width=0.65\textwidth]{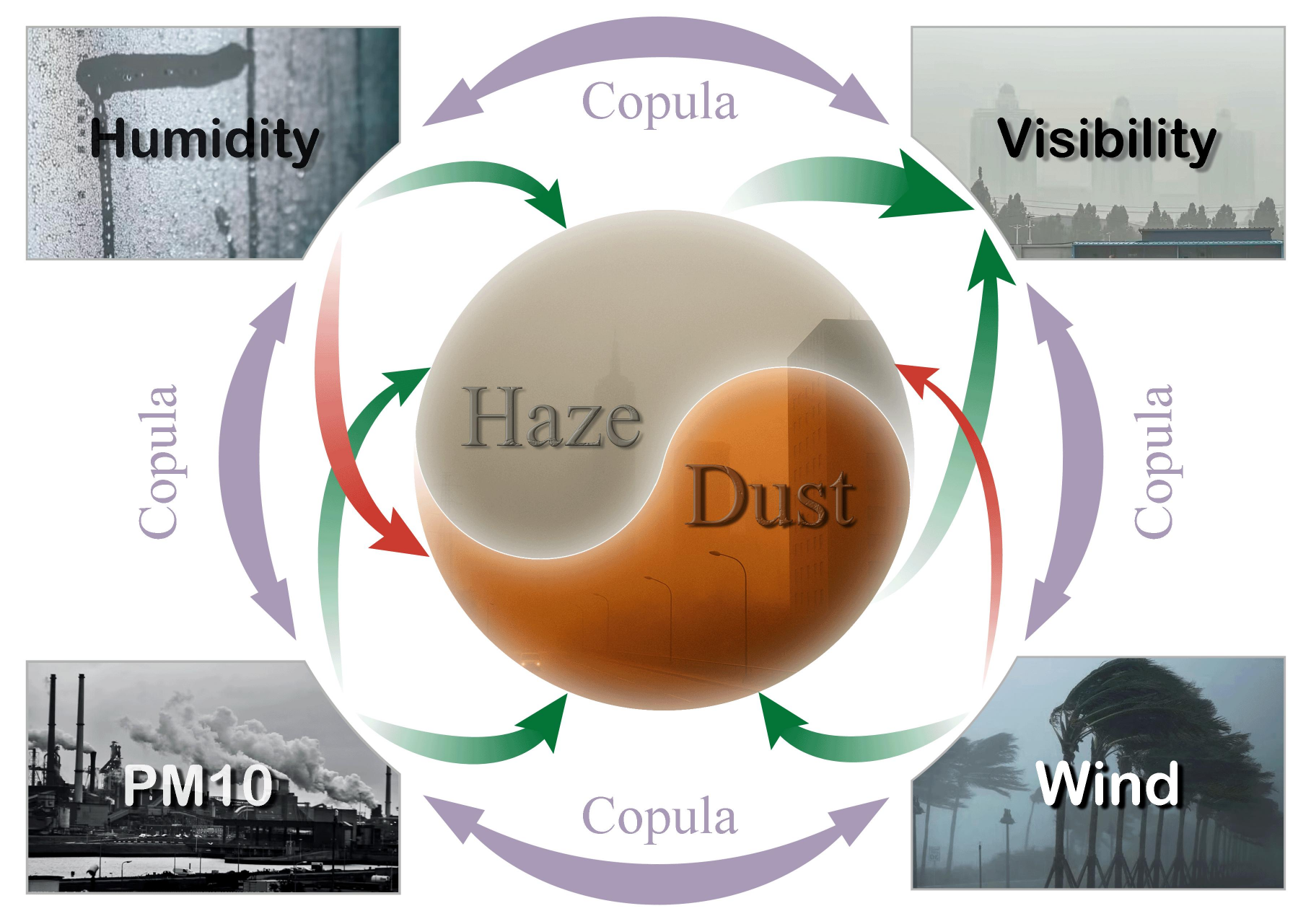}
    \caption{Interactions between observational indicators and pollution events. Green arrows denote positive effects and red arrows negative effects; arrow thickness encodes the mutual information (MI) with the event category, thereby reflecting the observational weight. The outer purple arrows indicate the use of a Gaussian copula to capture complex cross-dependence among indicators.}

    \label{fig:Logic Diagram of Observational Metrics}
\end{figure}

Importantly, \citet{Zhang2015HazeVisibility} observed that in Beijing, when relative humidity exceeds 80\%, aerosol hygroscopic growth alters the visibility-PM relationship: visibility can remain low even at moderate PM concentrations. This highlights the nonlinear and conditional dependencies among the chosen variables. To capture such complexities, this study models PM$_{10}$ concentration, wind speed, visibility, and relative humidity jointly using a Gaussian Copula, which allows flexible dependency structures without restricting correlations to linear forms. This approach provides both a physically meaningful and statistically robust basis for joint Haze-Dust classification. The conceptual relationships between the variables and pollution events are illustrated in Figure~\ref{fig:Logic Diagram of Observational Metrics}.

Meteorological data were downloaded from the Beijing Capital International Airport monitoring station, encompassing wind speed, visibility, relative humidity and categorical weather conditions (Clear, Haze, Dust). Historical air-quality records from the same site, including PM$_{10}$ concentrations and related variables, were likewise collected. Following source-wise imputation of missing values and deletion of a small number of irreparable gaps, a total of 8\,476 hourly observations covering 1~March~2023 -- 9~March~2024 were retained for subsequent analysis.

\subsubsection{Validation of emission distribution}
Accurately specifying the marginal distributions of each observed variable is a critical step when constructing the emission probabilities of the Factorial Hidden Markov Model (FHMM). The four key indicators selected in this study—\pmten concentration, mean wind speed, minimum visibility, and relative humidity—have their descriptive statistics summarized in the following table:

\begin{table}[H]
\centering
\caption{Descriptive Statistics: Key Meteorological and Pollution Indicators}
\label{tab:summary_stats}
\begin{tabular}{ccccccc}
\toprule
Variable & Mean & Std & Min & Max & Skewness & Kurtosis \\ 
\midrule
PM10 ($\mu g/m^{3}$) & 75.94 & 91.95 & 3 & 1897 & 7.01 & 84.21 \\
Average Wind Speed ($m/s$) & 2.83 & 2.18 & 0 & 15 & 1.64 & 2.88 \\
Minimum Visibility (km)\textsuperscript{a} & 9.25 & 1.81 & 1.1 & 10 & -2.45 & 4.96 \\
Relative Humidity (\%) & 54.42 & 26.06 & 5 & 100 & 0.09 & -1.16 \\
\bottomrule
\end{tabular}
\end{table}

Preliminary exploration reveals complex distributional features: All variables are non-negative; except for visibility (which exhibits truncation), the other indicators are right-skewed, ruling out direct use of a symmetric normal distribution.

\begin{figure}[H]
    \centering
    \includegraphics[width=0.9\textwidth]{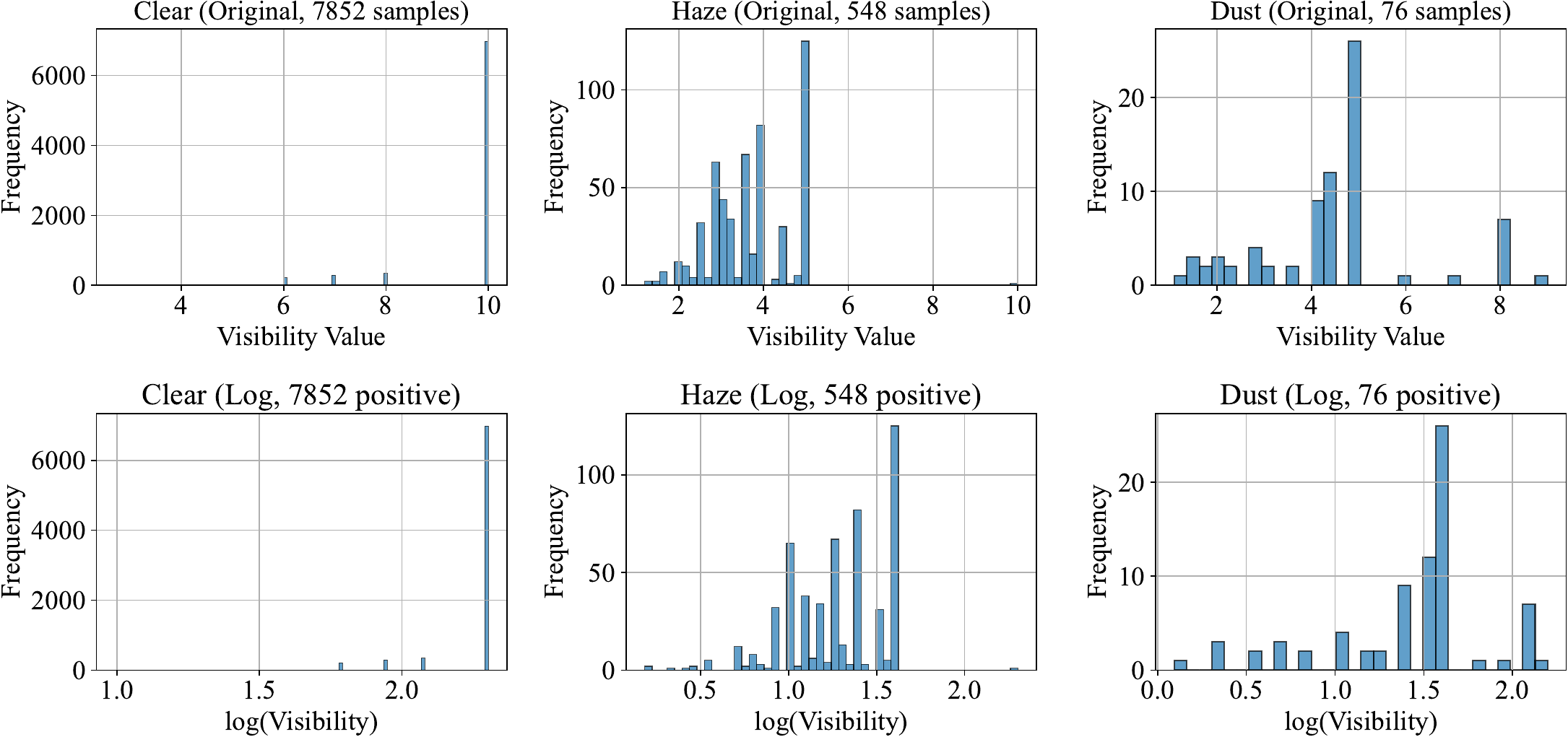}
    \caption{Right-Censored Inflation Characteristic of Visibility Data} 
    \label{fig:visibility_original_plot}
\end{figure}

For visibility, Figure~\ref{fig:visibility_original_plot} shows an inflated and right-censored pattern at 10km under “Clear” conditions, a feature not observed in the “Haze” and “Dust” states. A single continuous distribution cannot capture this behavior.

\begin{figure}[H]
    \centering
    \includegraphics[width=0.9\textwidth]{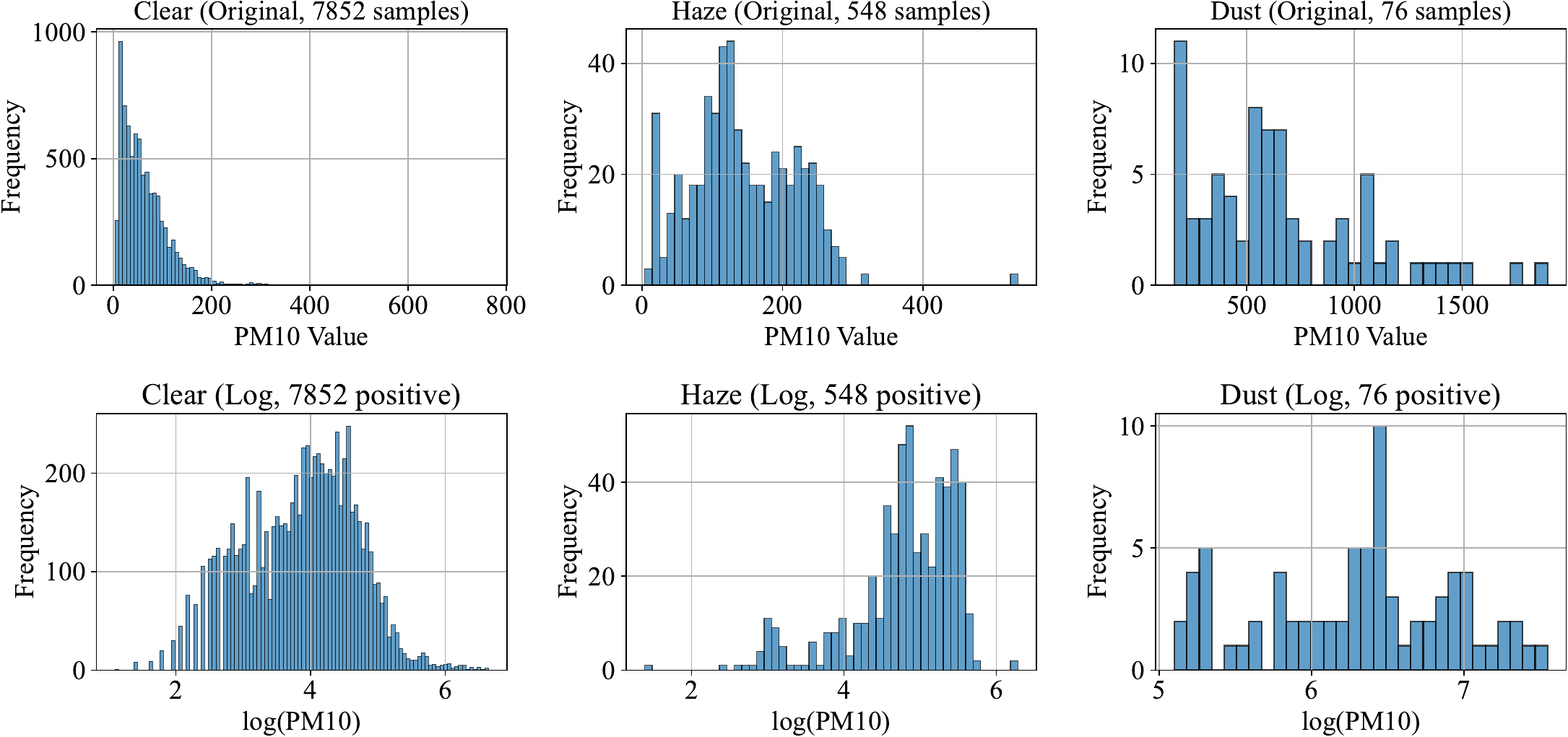}
    \caption{Distribution of \pmten Observations: Original vs. Log-Transformed} 
    \label{fig:\pmten _original_plot}
\end{figure}

For \pmten, wind speed, and relative humidity, we examine the effect of logarithmic transformation on their distributional shapes. Using \pmten as an example, Figure~\ref{fig:\pmten _original_plot} shows pronounced right-skewness across “Clear,” “Haze,” and “Dust” in the original scale; after log transformation, the distributions become much closer to symmetry and approximate normality across all three states. Similar trends are observed for wind speed and relative humidity.

\begin{figure}[H]
    \centering
    \includegraphics[width=\textwidth]{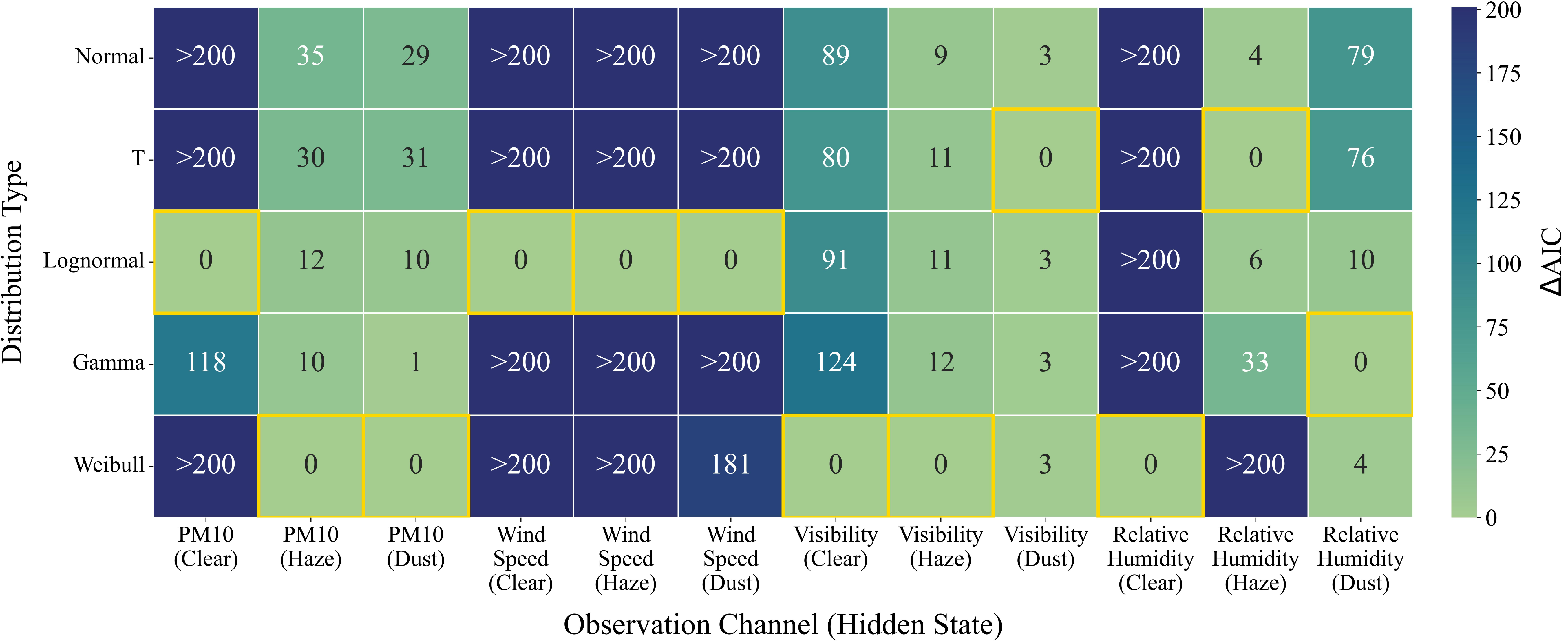} 
    \caption{$\Delta$AIC Heatmap: A comparison of goodness-of-fit for different probability distributions across various scenarios. The values represent the difference between each distribution's AIC and the optimal AIC for the current scenario; a smaller value indicates a better fit. Note that the data for Visibility (Clear) was pre-processed to account for its zero-inflation characteristic.}
    \label{fig:aic_heatmap}
\end{figure}

Motivated by these findings, the log-normal distribution is a strong candidate. To validate its suitability more rigorously, we compare several common distributions (including log-normal) using the Akaike Information Criterion (AIC). Figure~\ref{fig:aic_heatmap} presents a heatmap of $\Delta\text{AIC}=\text{AIC}-\text{AIC}_{\min}$ across scenarios. A value of $0$ indicates the best fit in a given scenario; values greater than $200$ are grouped and indicate substantially worse fits.

From the analysis of Figure \ref{fig:aic_heatmap}, we draw the following conclusions:

\begin{itemize}
    \item \textbf{No single universally optimal distribution:} No candidate dominates in all scenarios. For visibility, the $t$-distribution performs well by accommodating heavy tails; for \pmten under polluted states, the Weibull distribution shows a slight advantage.
    \item \textbf{Overall advantages and robustness of the log-normal:} Although not always achieving the lowest AIC, the log-normal exhibits consistently strong performance.
    \begin{itemize}
        \item In several important scenarios (e.g., all wind-speed cases), the log-normal is clearly optimal ($\Delta\text{AIC}=0$). In particular, for wind speed its AIC is substantially lower than alternatives, indicating excellent stability.
        
        \item Where it is not the best (e.g., \pmten in Haze/Dust; relative humidity in Dust), its $\Delta\text{AIC}$ typically remains small (about $10$), implying little loss relative to the optimum.
    \end{itemize}
\end{itemize}

Balancing these qualitative and quantitative results with model uniformity, parsimony, and numerical stability, we adopt the following emission assumptions:
\begin{itemize}
    \item \textbf{\pmten concentration, wind speed, relative humidity:} log-normal in all hidden states.
    \item \textbf{Visibility:} log-normal in “Haze” and “Dust”; inflated and truncated log-normal mixture in “Clear.”
\end{itemize}

\subsubsection{Parameter initialization}
For unlabeled datasets, we first apply K-means to the log-transformed observation vector
(\(\log\)PM$_{10}$, \(\log\mathrm{WS}\), \(\log\mathrm{RH}\), \(\log\mathrm{VIS})\)),
after removing inflated visibility points a priori. After clustering, samples at the inflated point are merged into the “no Haze-no Dust” state, and the emission parameters \((\mu,\sigma)\) for each hidden state are iteratively updated from the four clusters.

Because K-means cluster labels are unordered, we use a semi-supervised ordering strategy. Using prior means computed from historically labeled samples for the four typical weather combinations as priors, we compute each cluster’s mean observation vector and match it to the prior means by distance. Clusters are then ordered by increasing distance and mapped one-to-one to the model’s four hidden categories (Clear, Dust, Haze, and the joint “empty” state), thereby aligning cluster order with hidden states. Specific initialization parameters are provided in the Appendix.

\subsection{Model Configurations and Comparative Experiments}

To investigate two key design factors—(i) how to model dependence among observational variables and (ii) whether to incorporate mutual-information (MI) corrections in Viterbi decoding—we construct six model variants and compare them systematically. The specific configurations are summarised in Table~\ref{tab:ModCf}.

\begin{table}[H]
    \centering
    {\renewcommand{\arraystretch}{1.5}%
    \caption{Summary of Model Configurations}
    \label{tab:ModCf}
    \begin{tabular}{c c c c}
        \hline
        \textbf{Family / Model} & \textbf{Code} & \textbf{Viterbi Weighting} & \textbf{Reclassification} \\
        \hline
        Regular FHMM & M0  & None               & No  \\
        \hline
        \multirow{2}{*}{\makecell{Joint-Gaussian FHMM \\ (JG)}} 
                     & M1a & None               & \multirow{2}{*}{No}  \\
                     & M1b & Normalised MI      &                      \\
        \hline
        \multirow{3}{*}{\makecell{Log-Normal Copula FHMM \\ (LNC)}} 
                     & M2a & None               & \multirow{3}{*}{Yes} \\
                     & M2b & Raw MI             &                      \\
                     & M2c & Normalised MI      &                      \\
        \hline
        \multicolumn{4}{l}{\footnotesize
            Normalised MI:\,
            $\displaystyle 
              w_{i,(k_1,k_2)} = 
              \frac{I\bigl(x_{i};\, Z_{1}=k_1, Z_{2}=k_2\bigr)}
                   {\sum_{n=1}^{4} I\bigl(x_{n};\, Z_{1}=k_1, Z_{2}=k_2\bigr)}$
        } \\
    \end{tabular}}
\end{table}

In the “Viterbi Weighting” column we distinguish three variants: the regular Viterbi algorithm without weighting, Viterbi with mutual-information (MI) weights applied to the emission distribution, and Viterbi with \emph{normalised} MI weights.
For the Log-Normal Copula (LNC) family, simultaneous Haze-and-Dust events are extremely rare (fewer than 0.05 \% of all cases); these observations are reclassified a posteriori by a simple rule based on wind speed and relative humidity.
For the Joint-Gaussian (JG) models, any invalid assignments are merged into the Clear class.

\subsubsection{Micro-F1 and Macro-F1}

Given the pronounced class imbalance in our dataset (for example, the frequency of ‘Dust’ is less than 1\% of that of ‘Clear’), overall accuracy is not an informative metric. We therefore adopt the F1 score as the main criterion because it combines precision and recall. To provide a comprehensive view, we report both Micro-F1 and Macro-F1. The former reflects overall performance with sample-level aggregation, whereas the latter averages per-class performance to highlight balanced recognition of rare events such as ‘Dust’. This dual-metric scheme more fully reveals strengths and weaknesses when classifying imbalanced pollution events. The mathematical definitions of Micro-F1 and Macro-F1 are as follows\footnote{Precised confusion matrix see appendix.}:

Let there be $K$ classes (here $K=3$, namely Clear, Haze, and Dust). For class $k$, denote true positives, false positives, and false negatives by $\operatorname{TP}_k,\ \operatorname{FP}_k,\ \operatorname{FN}_k$. Then:
\begin{equation}
\text{Micro-F1}_k
=\frac{2 \sum_{k=1}^{K} \operatorname{TP}_k}
{2 \sum_{k=1}^{K} \operatorname{TP}_k+\sum_{k=1}^{K} \operatorname{FP}_k+\sum_{k=1}^{K} \operatorname{FN}_k}
\end{equation}
\begin{equation}
\text{Macro-F1}_k
=\frac{1}{K}\sum_{k=1}^{K}
\frac{2\operatorname{TP}_k}
{2\operatorname{TP}_k+\operatorname{FP}_k+\operatorname{FN}_k}
\end{equation}

Here, Micro-F1 first aggregates predictions across all samples and then computes F1; Macro-F1 first computes per-class F1 and then takes the arithmetic mean. They thus characterise overall and class-balanced performance, respectively. Micro-F1 emphasises global predictive accuracy and is suitable when samples are highly imbalanced, while Macro-F1 focuses on balanced recognition across pollution types.

\begin{figure}[H]
\centering
\includegraphics[width=1\textwidth]{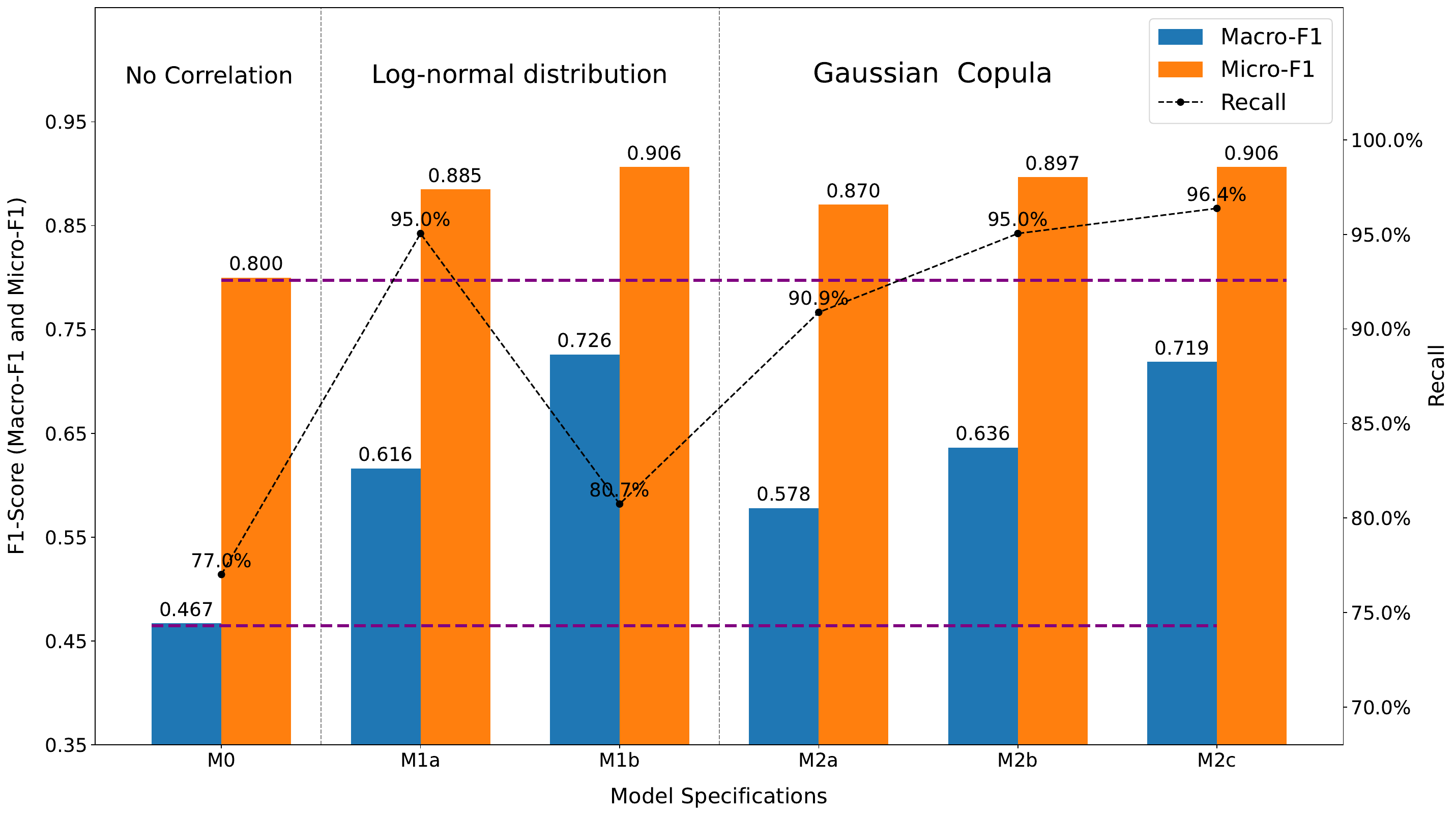}
\caption{Comparison of different models on F1 (Macro-F1 \& Micro-F1) and Recall (Mean of Haze \& Dust) metrics.}
\label{fig:f1_barplot}
\end{figure}

From Figure~\ref{fig:f1_barplot} and the detailed outputs, we observe the following:

M1a and M2a both introduce dependence but no weighting. Compared with the baseline M0 (no weighting, no dependence), both improve classification for all event types. Closer inspection shows M2a is more extreme: although it identifies all 76 Dust events and achieves high recall for rare classes (Haze \& Dust), the associated probability mass becomes overly concentrated, leading to a large number of false positives—an undesirable outcome that calls for further refinement.

Building on M1a, M1b adds normalised MI weights. It substantially improves Dust classification and yields more precise recognition of Clear; Macro-F1 rises to 0.726 (the highest in the table) and Micro-F1 reaches 0.906, indicating that weighting markedly enhances discriminative ability. While its overall F1 is the best and Haze recall is very high (548/548), Dust recall remains limited (47/76), representing a conservative profile of “low detection with low false alarms” for an extremely rare class.

Relative to M2a, M2b introduces MI weights without normalisation, increasing Macro-F1 to 0.636 and Micro-F1 to 0.897. The model still nearly recovers all Dust events (75/76); with weighting, Dust false positives are reduced by almost half, and Haze true/false positives both rise slightly, yielding a clear overall performance gain.

Normalising the weights in M2c makes the model more stable and precise. Macro-F1 improves to 0.719, and Micro-F1 ties M1b for the highest value (0.906). With stronger control over false positives, M2c attains the highest average recall for rare events within an acceptable false-positive range: Haze and Dust recalls reach 96.7\% and 98.7\%, respectively.

Overall, introducing weights—especially after normalisation—substantially enhances classification of rare categories, underscoring the value of information-theoretic priors. The joint log-normal structure tends to reduce false positives for rare events, whereas the Gaussian copula offers higher recall in rare-event detection. Below we report the final EM-estimated parameters used by M2a, M2b, and M2c; complete parameter sets and confusion matrices for all models are provided in the Appendix.

\begin{table}[htbp]
  \centering
  \small
  \caption{LNC Model: Shared Final Transition Matrices for M2a, M2b, and M2c}

  \subfloat[ Haze ]{
    \begin{tabular}{c|cc}
      \toprule
      \diagbox{from}{to} & 0 & 1 \\
      \midrule
      0 & 0.9887 & 0.0830 \\
      1 & 0.0113 & 0.9170 \\
      \bottomrule
    \end{tabular}
  }\qquad
  \subfloat[ Dust ]{
    \begin{tabular}{c|cc}
      \toprule
      \diagbox{from}{to} & 0 & 1 \\
      \midrule
      0 & 0.9937 & 0.0762 \\
      1 & 0.0063 & 0.9238 \\
      \bottomrule
    \end{tabular}
  }

\end{table}

\begin{table}[H]
    \centering
    \caption{Shared Final Mean ($\mu_{hk}$) and Standard Deviation ($\sigma_{hk}$) Parameter Matrices for Models M2a, M2b, and M2c}
    \label{tab:mu_sigma}
    
    % Use siunitx for decimal alignment
    % S[table-format=1.4] means 1 integer part, 4 decimal places
    \sisetup{
        table-number-alignment = center % Retain horizontal centering
    }
    
    \begin{tabular}{
        @{}          % No extra space on the left side of the table
        c            % Hidden State (k1, k2)
        c            % Statistic (μ, σ)
        S[table-format=1.4] % PM10 column
        S[table-format=1.4] % Wind column
        S[table-format=1.4] % RH column
        S[table-format=1.4] % Vis column
        @{}          % No extra space on the right side of the table
    }
        \toprule
        % \multicolumn is used to merge columns and center the header
        \multirow{2}{*}{Hidden State $(k_1,k_2)$} & 
        \multirow{2}{*}{Statistics} & 
        \multicolumn{4}{c}{Observation Variables} \\
        \cmidrule(l){3-6} % Use \cmidrule to draw a partial rule, (l) leaves a gap on the left
         & & {PM$_{10}$} & {Wind} & {Visibility} & {Humidity} \\
        \midrule
        
        \multirow{2}{*}{$(0,0)$} 
        & $\mu$    & 3.73 & 0.87 & 2.04 & 3.79 \\
        & $\sigma$ & 0.76 & 0.67 & 0.10 & 0.56 \\ 
        \addlinespace % Adds a little vertical space between row groups for better readability
        
        \multirow{2}{*}{$(0,1)$} 
        & $\mu$    & 5.33 & 0.96 & 1.90 & 3.53 \\
        & $\sigma$ & 0.60 & 0.72 & 0.36 & 0.72 \\
        \addlinespace
        
        \multirow{2}{*}{$(1,0)$} 
        & $\mu$    & 4.50 & 0.49 & 1.65 & 4.45 \\
        & $\sigma$ & 0.67 & 0.49 & 0.44 & 0.13 \\
        
        \bottomrule
    \end{tabular}
\end{table}

As mentioned in the previous chapter, for the Gaussian Copula model, we use a global correlation matrix that is independent of the hidden states, aiming to capture the physical relationships between the observational variables.

\begin{figure}[H]
\centering
\includegraphics[width=0.6\textwidth]{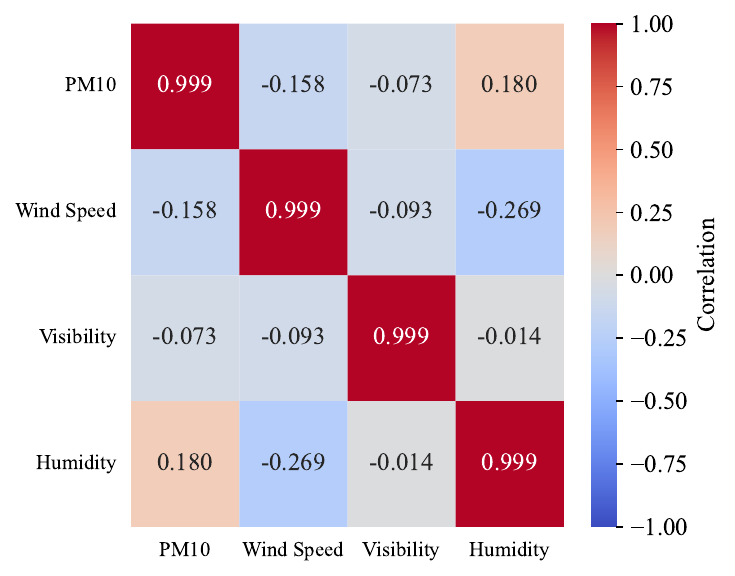}
\caption{Heatmap of the final global correlation matrix for models M2a, M2b, and M2c}
\label{fig:corr_heatmap_M2c}
\end{figure}

\subsubsection{ROC and AUC}

To further quantify performance differences between the joint log-normal and Gaussian-copula emission models in multi-class discrimination, we also consider classification by maximum a posteriori (MAP) using posterior probabilities. Comparing the two ROC curves and their AUCs provides an intuitive assessment of decision-boundary quality and overall discriminative ability under different pollution states.

\begin{figure}[H]
  \centering
  \begin{subfigure}[b]{0.46\textwidth}
    \centering
    \includegraphics[width=\textwidth]{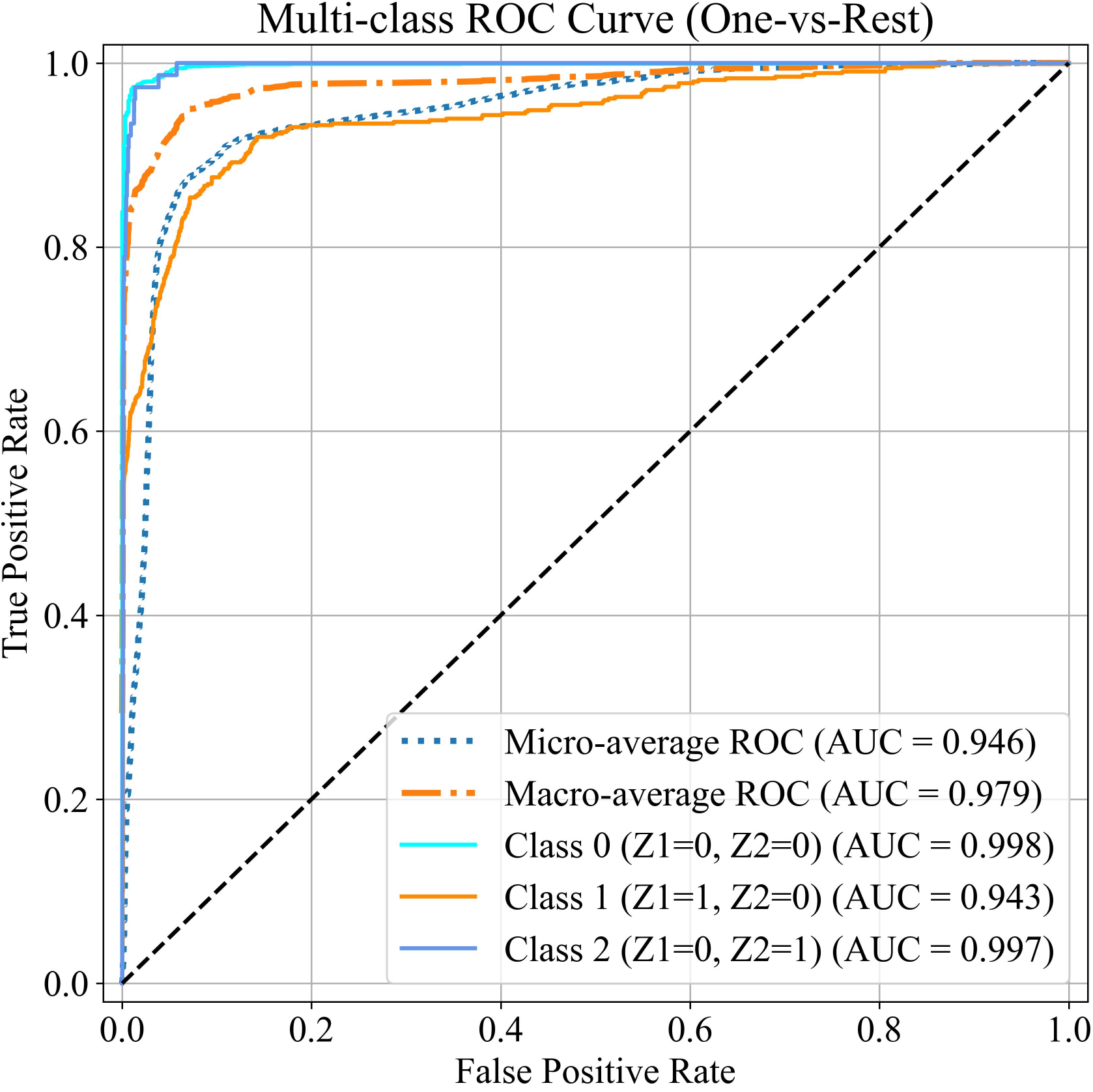}
    \caption*{(a) Joint Log-normal Distribution}
  \end{subfigure}
  \hfill
  \begin{subfigure}[b]{0.46\textwidth}
    \centering
    \includegraphics[width=\textwidth]{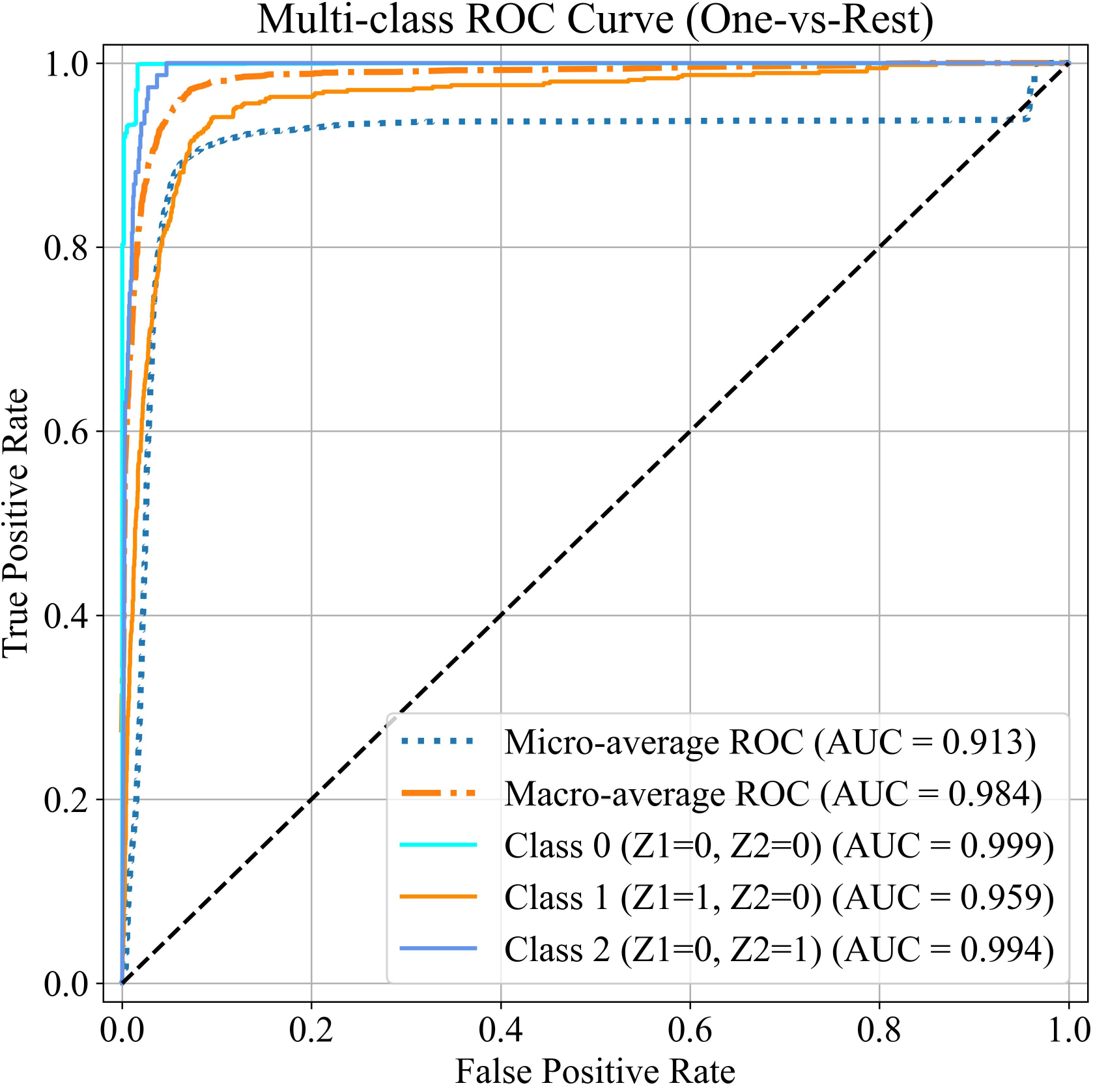}
    \caption*{(b) Gaussian Copula}
  \end{subfigure}
  \caption{Multi-class ROC curves for the two emission models: Joint Log-normal Distribution and Gaussian Copula.}
  \label{fig:roc_comparison}
\end{figure}

The ROC/AUC comparison shows that both methods achieve nearly perfect separation for Clear and Dust (Class 0/2 AUC $>0.99$). For the extremely rare Dust class, the Gaussian copula attains a higher one-vs-rest AUC (0.997 vs. 0.994), indicating that explicitly modelling (log-)dependence yields more concentrated and stable probability ranking under extreme joint features (high PM$_{10}$, high wind speed, low humidity, and low visibility). In contrast, the Haze AUC is slightly lower (0.943 vs. 0.959), producing a modest drop in macro-averaged AUC (0.979 vs. 0.984), which suggests some loss of fine-grained resolution for boundary-like, moderate-pollution cases. However, on the micro-averaged AUC—which emphasises global ranking and calibration—the Gaussian copula clearly outperforms the log-normal baseline (0.946 vs. 0.913), pointing to improved overall score discrimination under class imbalance. In sum, the two approaches perform similarly for common conditions (Clear), while their differences on rare extremes more clearly reflect sensitivity to joint-tail structure: the Gaussian copula maintains very high detection of extreme events and improves global probability ranking, whereas the joint log-normal distribution retains slightly better local separability for Haze.

\subsection{Mutual Information-Derived Weight Effect}

\begin{figure}[H]
  \centering
  %-- Two plots in the first row --%
  \begin{subfigure}[t]{0.5\textwidth}
    \centering
    \includegraphics[width=\linewidth]{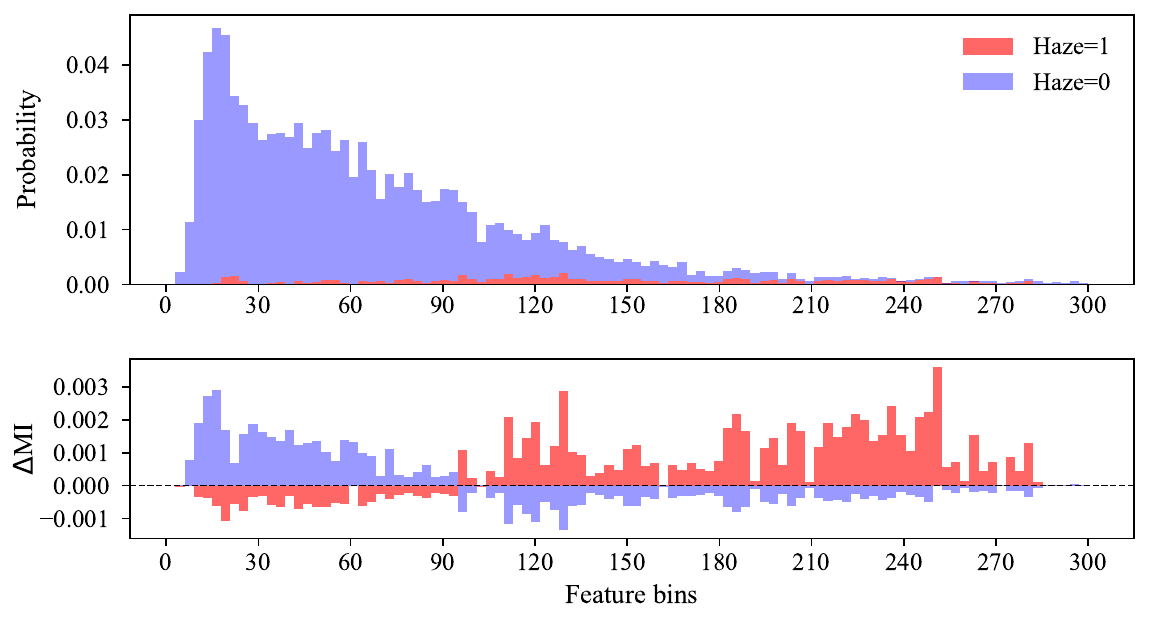}
    \caption*{(a) \(PM_{10}\)}
  \end{subfigure}%
  \hfill
  \begin{subfigure}[t]{0.5\textwidth}
    \centering
    \includegraphics[width=\linewidth]{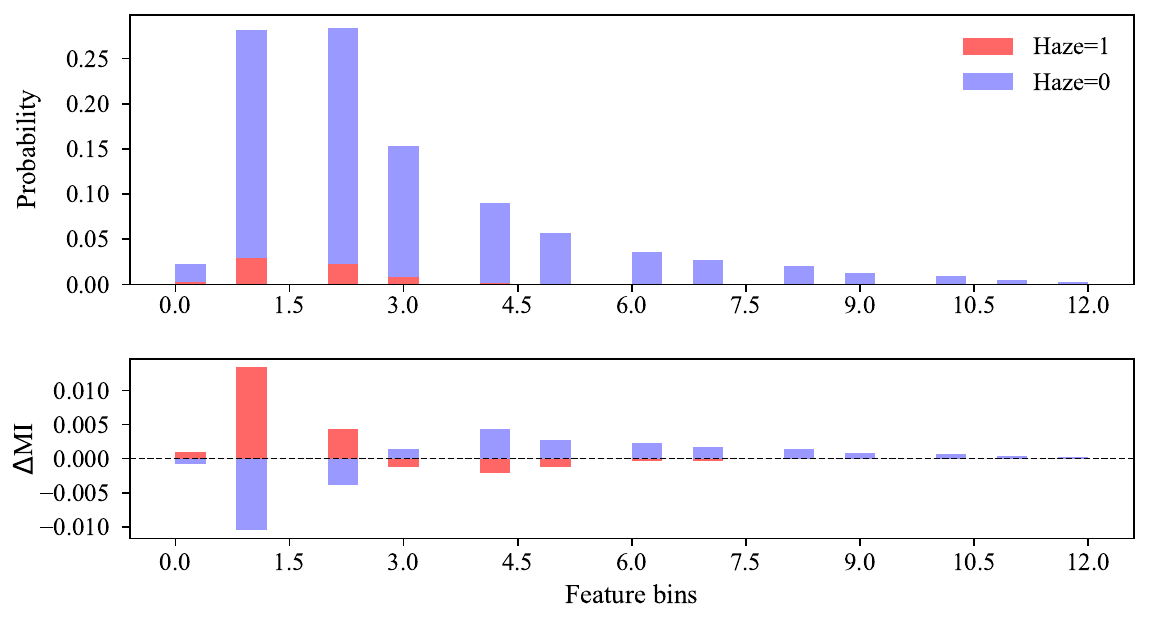}
    \caption*{(b) Wind Speed}
  \end{subfigure}

  \medskip

  %-- One plot in the second row --%
  \begin{subfigure}[t]{0.5\textwidth}
    \centering
    \includegraphics[width=\linewidth]{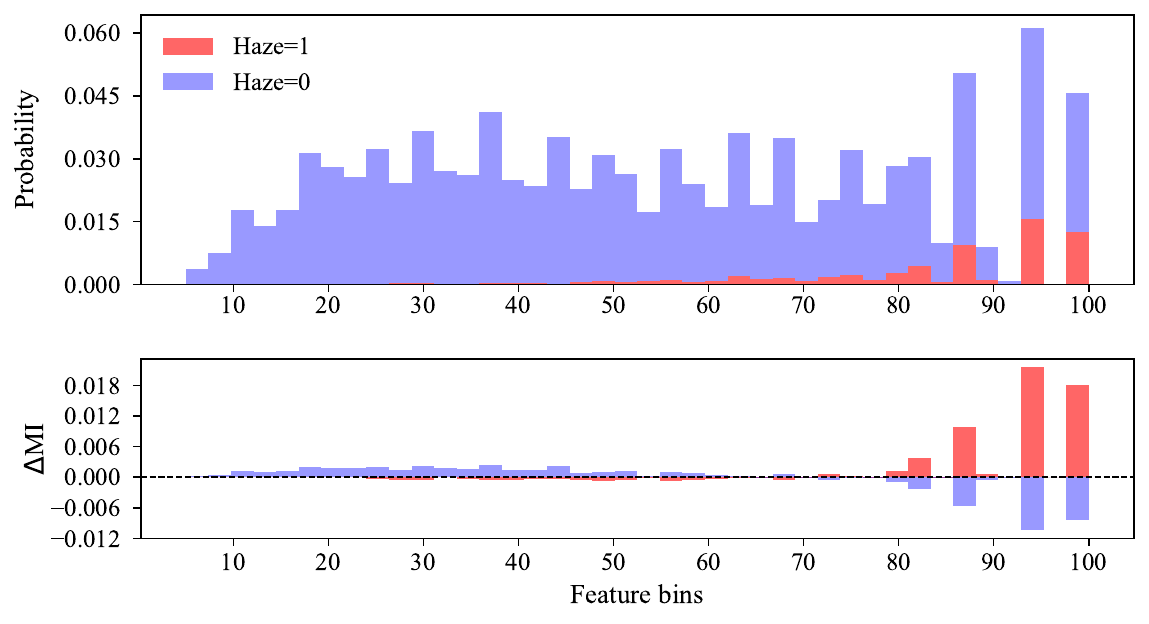}
    \caption*{(c) Relative Humidity}
  \end{subfigure}

  %-- Caption, using \parbox to auto-wrap the formula --%
  \caption{%
    Example of Mutual Information decomposition: Probability distributions of three observational features for discriminating Haze (top) and their local mutual information contribution:
    \parbox{\textwidth}{%
      \[
      \Delta I_{\boldsymbol x,Z^{\mathrm{Haze}}}
      =\Pr(\boldsymbol x,Z^{\mathrm{Haze}})\,
         \log\frac{\Pr(\boldsymbol x,Z^{\mathrm{Haze}})}
           {\Pr(\boldsymbol x)\,\Pr(Z^{\mathrm{Haze}})}.
      \]
    }%
  }
  \label{fig:mi_decompose_three}
\end{figure}

Mutual information $I(\boldsymbol x;\boldsymbol Z)$ is a global scalar that quantifies how much the uncertainty in $\boldsymbol Z$ is reduced once the feature vector $\boldsymbol x$ is observed. Table~\ref{tab:viterbi_weights_single} lists the normalised weights for the four observational variables derived from mutual information, and Figure~\ref{fig:mi_decompose_three} further illustrates where these weights originate.

\begin{table}[H]
  \centering
  \small
  \caption{Normalized Weights for Observational Metrics per Pollution Event in the Viterbi Stage}
  \label{tab:viterbi_weights_single}

  \begin{tabular}{c|cccc}
    \toprule
    \diagbox{Pollution Event}{Observation} & PM10   & Wind Speed &  Visibility   & Humidity \\
    \midrule
    Clear Event      & 0.2324 & 0.1957     & 0.3443        & 0.2276 \\
    Dust Event       & 0.2535 & 0.2477     & 0.2500        & 0.2488 \\
    \midrule
    Haze Event       & 0.2327 & 0.1979     & 0.3408        & 0.2285 \\
    Haze \& Dust Event & 0.2500 & 0.2500     & 0.2500        & 0.2500 \\
    \bottomrule
  \end{tabular}
\end{table}

%------------------------------------------------------------
Taking Haze as an example, we illustrate how mutual information is computed for the joint effects of \pmten, wind speed, and relative humidity. Mutual information can be written as
\[
I(\boldsymbol x;Z^{\text{Haze}})=\sum_{\boldsymbol x}\sum_{Z^{\text{Haze}}}\Pr(\boldsymbol x,Z^{\text{Haze}})\,
\log\frac{\Pr(\boldsymbol x,Z^{\text{Haze}})}{\Pr(\boldsymbol x)\,\Pr(Z^{\text{Haze}})},
\]
where $\Pr(\boldsymbol x,Z^{\text{Haze}})$ is the joint probability of $\boldsymbol x$ and $Z^{\text{Haze}}$, and $\Pr(\boldsymbol x)$ and $\Pr(Z^{\text{Haze}})$ are the respective marginals.
Define
\[
\Delta I_{\boldsymbol x,Z^{\text{Haze}}}=\Pr(\boldsymbol x,Z^{\text{Haze}})\,
\log\frac{\Pr(\boldsymbol x,Z^{\text{Haze}})}{\Pr(\boldsymbol x)\,\Pr(Z^{\text{Haze}})},
\]
so that $\Delta I_{\boldsymbol x,Z^{\text{Haze}}}$ describes the local contribution of the value pair $(\boldsymbol x,Z^{\text{Haze}})$ to the overall mutual information.
The larger the deviation between $\Pr(\boldsymbol x,Z^{\text{Haze}})$ and $\Pr(\boldsymbol x)\Pr(Z^{\text{Haze}})$ under independence, the larger $\lvert\Delta I_{\boldsymbol x,Z^{\text{Haze}}}\rvert$; if they are equal, then $\Delta I_{\boldsymbol x,Z^{\text{Haze}}}=0$.
Summing all $\Delta I_{\boldsymbol x,Z^{\text{Haze}}}$ yields the global $I(\boldsymbol x;Z^{\text{Haze}})$.
Hence, the distribution of $\{\Delta I_{\boldsymbol x,Z^{\text{Haze}}}\}$ reveals the main source intervals of mutual information and the strength of dependence between feature values and the target class.

From Figure~\ref{fig:mi_decompose_three}, we draw the following conclusions:
\begin{itemize}
  \item \textbf{\pmten:} Total information $I\!=0.2327$. The high-concentration tail ($\text{\pmten}>180\,\mu\mathrm{g/m}^{3}$) contains dense Haze samples and contributes the main positive $\Delta I$, indicating that \pmten is a key “pollution-intensity” feature.
  \item \textbf{Wind speed:} $I\!=0.1979$. Under calm to light winds ($<4\,\mathrm{m/s}$), the Haze probability increases markedly; under strong winds ($>7\,\mathrm{m\,s}^{-1}$), Haze and wind are largely mutually exclusive, yielding negative $\Delta I$.
  \item \textbf{Relative humidity:} $I\!=0.2285$. The high-humidity regime ($>80\%$) contributes most to Haze discrimination; the mid-humidity range carries almost no information; low humidity contributes negative bars, suggesting that humidity mainly enhances optical extinction by promoting secondary aerosol formation.
\end{itemize}

In sum, the three variables have comparable total mutual information (about $0.20$--$0.23$), but their informative ranges are clearly complementary. Specifically, the positive contributions of \pmten and RH concentrate in their upper tails (high concentration / high humidity), reflecting particle accumulation and humidity-enhanced optical extinction during Haze; by contrast, Wind contributes positively in its lower tail (calm to light winds), consistent with the increased likelihood of Haze under stable conditions, while at higher wind speeds its contribution approaches zero or turns negative. These results corroborate our physics-based variable selection and imply a clear modelling guideline: to enhance Haze identification, emphasise the co-occurrence pattern “high \pmten + high RH + low Wind.” In addition, because the local term $\Delta I$ is logarithmic in nature, it is particularly sensitive to joint tail events; prior comparisons also indicate that a Gaussian copula alone has limited ability to capture tail dependence among indicators under rare events. Therefore, combining MI-derived weights or priors with the Gaussian copula model can partially compensate for this shortcoming without materially increasing model complexity, thereby improving robustness and interpretability.

\subsection{Sensitivity of Weight Normalization and Global Weight}
\label{sec:weight_sensitivity}

% --- Part 1: Motivation and Method ---
\subsubsection{Motivation and Method: Introducing a Global Weight \textit{v}}

The Viterbi algorithm seeks the hidden-state sequence that maximises the product of emission and transition probabilities. Since we previously assigned per-dimension weights based on mutual information and obtained their normalised sum \(\Sigma w\), it is likewise desirable to introduce a global control to adjust the relative share of the emission term in the objective. To this end, we introduce a global weight \(v>0\) as a hyperparameter that systematically balances the influence of observational data (emission probabilities) and model dynamics (transition probabilities).

This parameter acts by modulating the “sharpness” of the emission distribution:

When $v > 1$, the distribution flattens, reducing the model’s “confidence” in the current observation and relatively increasing the weight of transition history.

When $v < 1$, the distribution becomes sharper, strengthening dependence on the current observation and relatively increasing the weight of the emission probability.

For example, applying $v$ to a standard normal \(\mathcal{N}(0,1^2)\) changes its standard deviation to \(\sqrt{v}\): when $v=4$ it becomes $2$ (flattened), and when $v=0.25$ it becomes $0.5$ (sharpened).

As with the weighting above, raising a normalised probability density function to the power \(1/v\) breaks its probabilistic properties, so re-normalisation is required. In principle, this is achieved by integrating over the observation space to ensure a valid probability distribution:
\begin{equation}
    b_{(k_1,k_2)}^{\langle w,v\rangle}(\boldsymbol x_t) = \pr^{\text{Global Wtd}}(\boldsymbol x_t \mid \boldsymbol Z_t = \boldsymbol Z)= \frac{\left[  \pr^{\text{Weighted}}(\boldsymbol {x}_t \mid \boldsymbol Z_t)  \right] ^{1/v}}{C^{\text{Global Wtd}}_{(k_1,k_2)}} 
\end{equation}
where
\begin{equation}
    C^{\text{Global Wtd}}_{(k_1,k_2)}(v)=\idotsint_{\mathbb{R}^{E}} \left[  \pr^{\text{Weighted}}(\boldsymbol {x}_t \mid \boldsymbol Z_t = \boldsymbol Z) \right] ^{1/v}\,d\boldsymbol x
\end{equation} 

In practice, unlike the approximation used in~\eqref{eq:emisAdj}, we adopt an algorithmically equivalent and computationally feasible alternative: at each time step we normalise by the sum of (unnormalised) scores over all hidden states,
\begin{equation}
  \widetilde{C}^{\text{Global Wtd}}(v, \boldsymbol {x}_t) := \sum_{\boldsymbol Z_t \in \mathcal{Z}} N_{\boldsymbol Z_t}(v, \bm{x}_t) = \sum_{\boldsymbol Z \in \mathcal{Z}}\left[\pr^{\text{Weighted}}(\boldsymbol {x}_t \mid \boldsymbol Z_t = \boldsymbol Z) \right] ^{1/v}
\end{equation}
where $\widetilde{C}^{\text{Global Wtd}}$ depends on the observation $\boldsymbol {x}_t$ and serves as a surrogate for the intractable \(C^{\text{Global Wtd}}_{k_1,k_2}\). Substituting it into~\eqref{eq:temperature} yields \(\pr^{\text{Global Wtd}}(\boldsymbol x_t \mid \boldsymbol Z_t)\). The posterior can then be computed exactly as:
\begin{equation}
  \Pr(\boldsymbol Z_t = \boldsymbol Z_t \mid \bm{x}_t) = \frac{N_Z}{\widetilde{C}^{\text{Global Wtd}}}
\end{equation}
A detailed derivation is provided in Appendix~\ref{subsec:global-temp-norm}.

We refer to the model with both per-observation weights $w$ and the global weight $v$ as the doubly weighted model. Its Viterbi recursion becomes:
\begin{equation}
    \omega_t(j) = \max_i \biggl[ \omega_{t-1}(i) + \log\pr(Z_t=j\mid Z_{t-1}=i) \biggr] + \log\pr^{\text{Global Wtd}}(x_t\mid Z_t=j)
\label{eq:viterbi_update}
\end{equation}
where $i, j$ denote the hidden states at times $t-1$ and $t$, respectively.

% --- Part 2: Parameter Sensitivity Experiment ---
\subsubsection{Parameter Sensitivity Experiment}

In the preceding analysis, the normalization constant for the MI-based weights, $\Omega$, was fixed at 1. In practice, both $\Omega$ and the global weight $v$ jointly control the “sharpness” of the emission distribution. To examine their combined effects, we set $\Omega = \Sigma w$ (the row-sum of normalized weights) and conduct a two-dimensional sensitivity analysis.

\begin{figure}[H]
    \centering
    \subfloat[Overlay of Macro \& Micro F1]{%
    \includegraphics[width=1\textwidth]{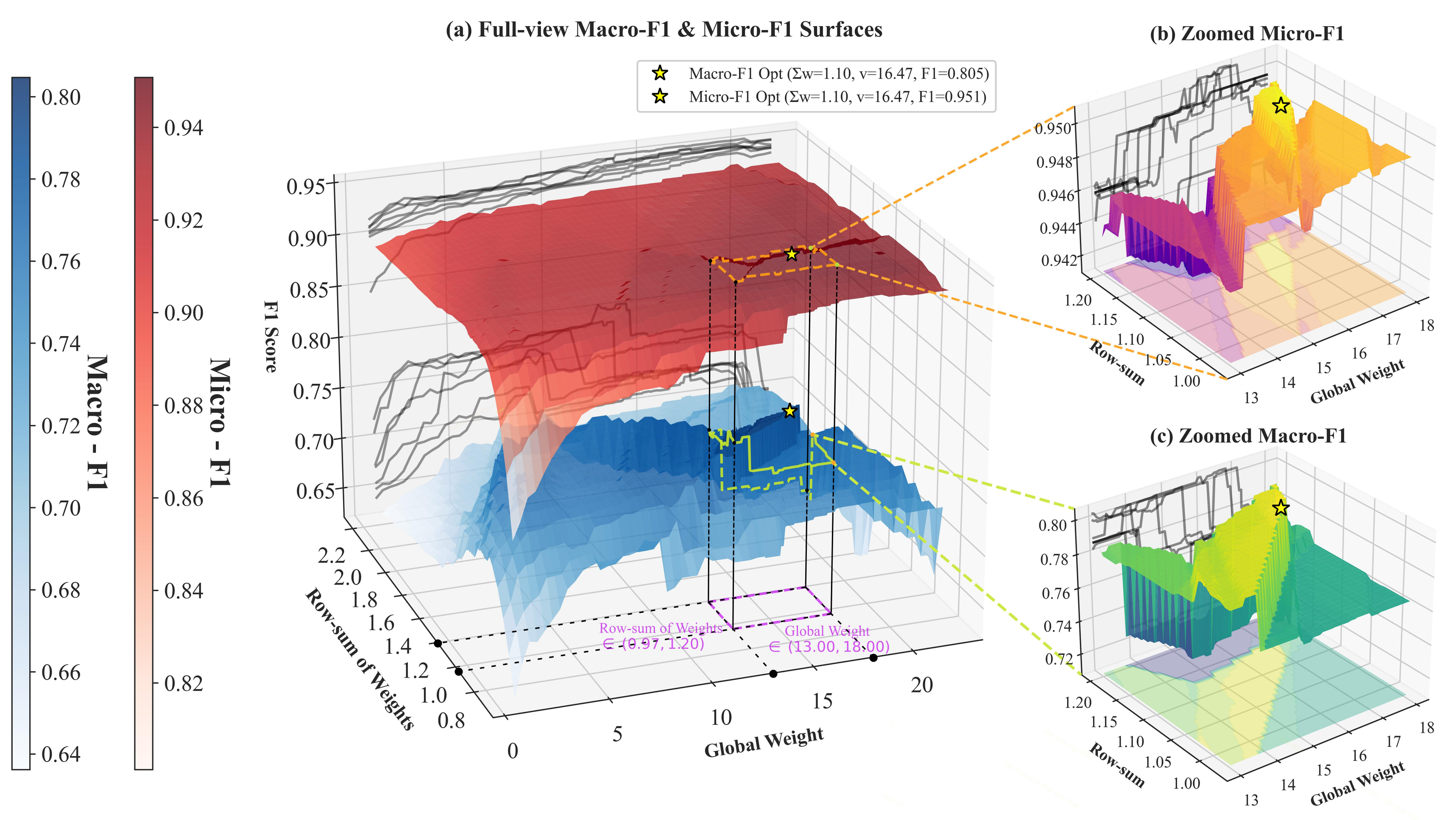}
    }
    \caption{Parameter landscapes obtained by jointly varying
           the row-sum of normalized weights (\(\Sigma w\))
           and the global weight (\(v\)). Starred markers indicate the
           optimal points on each surface.}
    \label{fig:overlay_landscape}
\end{figure}

Using the best-performing Gaussian-copula model (M2c) as the baseline, we perform a grid search over \(\Sigma w \in [0.7,\, 2.2]\) and \(v \in [0.7,\, 22.0]\), recording both Macro-F1 and Micro-F1. We then refine the grid in the peak region \((\Sigma w \in [0.97,\, 1.20],\ v \in [13.0,\, 18.0])\) for higher resolution. The resulting parameter landscapes are shown in Figure~\ref{fig:overlay_landscape}.

The 3D performance surfaces in Figure~\ref{fig:overlay_landscape} reveal that both Micro-F1 and Macro-F1 attain their maxima when \(\Sigma w \approx 1.10\) and \(v \approx 16.47\). Within the rectangular neighborhood defined by \(\Sigma w \in [1.04,\, 1.15]\) and \(v \in [15.73,\, 16.83]\), all sampled points achieve Micro-F1 scores of at least \(0.941\) and Macro-F1 scores of at least \(0.725\), indicating strong robustness around the optimum.

A notable finding is that optimal performance occurs at a relatively large \(v\), implying that the model benefits from emphasizing temporal dependencies (transition probabilities) over instantaneous observations (emission probabilities). This adjustment helps counteract potential over-reliance on single-time-step evidence and fully exploits the structural advantage of the \textsc{FHMM} in capturing temporal dynamics. From a practical perspective, the tunable \((\Sigma w, v)\) combination offers a flexible mechanism for balancing observational evidence against temporal priors—allowing practitioners, subject to application-specific constraints such as minimum recall, to target performance according to operational needs.

% --- Part 3: Final Model Performance Evaluation ---\subsubsection{Performance Evaluation of the Optimal Model}

Based on the parameter sensitivity analysis, we select a configuration that jointly optimizes Macro-F1 and Micro-F1 to train the final doubly weighted model, denoted M2d. Its full confusion matrix is provided in the Appendix. Table~\ref{tab:model_comparison_professional} presents a detailed comparison of M2d against the baseline (M0), the copula-enhanced model (M2a), and the MI-weighted model (M2c).

\begin{table}[H]
    \centering
    \footnotesize
    \caption{Performance comparison of the baseline, weighted, and parameter-optimized models. 
    M0: baseline with no correlation structure or weights; 
    M2a: Gaussian Copula correlation structure; 
    M2c: Copula + mutual information (MI) weights; 
    M2d: final model with optimized $\Sigma w$ and $v$.}
    \label{tab:model_comparison_professional}
    \begin{tabular*}{\textwidth}{@{\extracolsep{\fill}} l cccc ccc}
        \toprule
        \multirow{2}{*}{Metric} & \multicolumn{4}{c}{Model Performance} & \multicolumn{3}{c}{Relative Improvement (\%)} \\
        \cmidrule(lr){2-5} \cmidrule(lr){6-8}
        & M0 & M2a & M2c & M2d & vs. M0 & vs. M2a & vs. M2c \\
        \midrule
        Macro-F1 & 0.4671 & 0.5779 & 0.7188 & \textbf{0.8016} & +71.6 & +38.7 & +11.5 \\
        Micro-F1 & 0.7999 & 0.8702 & 0.9064 & \textbf{0.9490} & +18.6 & +9.1  & +4.7  \\
        Macro AUC          & 0.9336 & 0.9800 & 0.9800 & \textbf{0.9800} & +5.0  & —       & —       \\
        Micro AUC          & 0.9131 & 0.9459 & 0.9459 & \textbf{0.9459} & +3.6  & —       & —       \\
        \midrule
        \multicolumn{8}{l}{\textit{Per-Class F1 Scores}} \\
        \midrule
        Clear (Class 0) & 0.8981 & 0.9312 & 0.9478 & \textbf{0.9723} & +8.3   & +4.4  & +2.6  \\
        Haze (Class 1) & 0.3179 & 0.5918 & 0.5979 & \textbf{0.6839} & +115.1 & +15.6 & +14.4 \\
        Dust (Class 2) & 0.1854 & 0.2108 & 0.6109 & \textbf{0.7485} & +303.7 & +255.2& +22.5 \\
        \bottomrule
    \end{tabular*}
\end{table}

As shown in Table~\ref{tab:model_comparison_professional}, the optimized M2d model consistently outperforms all other variants across key metrics. Relative to the baseline M0, Macro-F1 rises from 0.4671 to \textbf{0.8016} (+71.6\%), while Micro-F1 increases from 0.7999 to \textbf{0.9490} (+18.6\%). AUC values—derived from EM posteriors—are identical for M2a, M2c, and M2d because all employ the copula structure during EM and differ only in the Viterbi decoding stage; all exceed the M0 baseline.

The most substantial gains occur for the rare classes. For “Dust,” the F1 score improves from 0.1854 to \textbf{0.7485} (+\textbf{303.7\%}); for “Haze,” it rises from 0.3179 to \textbf{0.6839} (+115.1\%). Compared with M2c, M2d delivers an additional 11.5\% Macro-F1 gain and a 22.5\% increase in Dust-class F1; compared with M2a, these gains are 38.7\% and 255.2\%, respectively.

Overall, the joint optimization of $\Sigma w$ and $v$ not only mitigates class imbalance but also improves generalization and stability. By slightly reducing rare-event recall while increasing recall for the majority “Clear” class, M2d lowers the false alarm rate (FAR) across all categories, achieving a more favorable precision-recall trade-off and delivering substantial F1 score improvements.

% ------------------------ %
% ------------------------ %
\section{Conclusion}

This study tackles the joint classification of two prevalent atmospheric pollution events—Haze and Dust—by introducing a novel Factorial Hidden Markov Model (FHMM)-based framework. By assuming independence among hidden chains and integrating fast inference strategies, the proposed framework addresses the computational bottlenecks encountered by conventional HMMs in high-dimensional settings, offering a practical and scalable approach for dynamic environmental system modeling. The combined use of chain independence and optimization techniques such as the Hadamard-Walsh transform substantially reduces both computational and memory requirements, thereby enabling operational deployment in high-temporal-resolution environmental applications.

At the dependency modeling level, the Gaussian Copula-based Log-Normal (GCLN) model demonstrably outperforms the traditional multivariate log-normal approach in maintaining balanced classification performance across rare event categories. Empirically, the recall for Haze and Dust reaches 96.7\% and 98.7\%, respectively. Incorporating mutual information-based weighting further enhances rare-class recognition under imbalanced conditions. In quantitative terms, the best-performing model (M2c) achieves a Micro-F1 score of 0.9064 and a Macro-F1 score of 0.7188. A stepwise performance trajectory confirms the contribution of each methodological enhancement: from the baseline (Micro-F1\(_{M0} = 0.7999\)) to the Copula-augmented model (Micro-F1\(_{M2a} = 0.8702\)), and ultimately to the mutual-information-weighted variant (Micro-F1\(_{M2c} = 0.9064\)).

Building upon these improvements, the Viterbi decoding algorithm was reformulated to maximize the product of transition and weighted emission probabilities. A tunable balancing mechanism was introduced to control the relative influence of observational evidence versus transition dynamics. Specifically, the design combines mutual information-derived per-variable weights (\(\Sigma w\)) with a global static weight \(v\), enabling fine-grained calibration of emission “sharpness” under imbalanced class distributions. This statistically grounded approach yields substantial performance gains: Micro-F1 increases to 0.9490, Macro-F1 to 0.8016, with F1 scores for Haze and Dust improving by 14.4\% and 22.5\%, respectively.

The methodological principles advanced here—copula-based dependency modeling, mutual information-guided feature weighting, and hyperparameter optimization for rare-event sensitivity—are broadly transferable within atmospheric science. Potential applications include compound pollution detection, cross-scale pollutant transport attribution, early warning systems for extreme weather-pollution coupling, and multi-source observational data fusion. Consequently, this work not only introduces an effective classification tool for Haze and Dust events but also contributes to the methodological foundations for air quality risk assessment and decision support.

While the proposed framework demonstrates strong empirical performance, several limitations invite further research. Notably, the assumption of hidden chain independence constitutes a simplification of physical reality. Although this assumption ensures tractability, future work should consider relaxing or refining it. We identify two primary directions:

\begin{enumerate}
    \item \textbf{Model Generalization and Extension:} Extend the framework to larger geographic domains and more complex pollution phenomena (e.g., ozone-particulate co-pollution), incorporate seasonal dynamics into the temporal model, and increase the number of hidden chains to capture a richer spectrum of event types beyond binary categorization.
    \item \textbf{Dynamic Weighting Mechanisms:} Develop adaptive weighting strategies that respond to real-time meteorological conditions or event stages, thereby improving responsiveness to abrupt or extreme pollution episodes.
\end{enumerate}

% ------------------------ %
% ------------------------ %
\section*{Appendix}
\addcontentsline{toc}{section}{Appendix}
\renewcommand{\thesubsection}{\Roman{subsection}}
\setcounter{subsection}{0} 

% ------------------------------------------------------------
\subsection{Proof of the suitability of the Gaussian copula for the log-normal distribution}
Using a Gaussian copula to model the dependence structure of a multivariate log-normal distribution is fully equivalent to adopting the covariance matrix of the joint log-normal distribution itself, thereby capturing the correlation exactly, and we need only show that, when the true distribution is joint log-normal, the left- and right-hand sides of the equation

\[
\widetilde b_{(k_{1},k_{2})}(x_{t})
=|\boldsymbol  R_{(k_{1},k_{2})}|^{-\tfrac12}
\exp\left\{-\tfrac12\boldsymbol  Z_{t}^{\top}\bigl(\boldsymbol  R_{(k_{1},k_{2})}^{-1}-\boldsymbol  I\bigr)\boldsymbol  Z_{t}\right\}
\prod_{i=1}^{p}\mathcal{LN}\!\bigl(x_{t,i};\mu^{(k_{1},k_{2})}_{i},\sigma^{2\,(k_{1},k_{2})}_{i}\bigr),
\]
are identical.

\paragraph{Step 1—Write out the joint density of the log-normal distribution}

Let
\[
\mathbf x=(x_1,\dots,x_E)\sim\mathcal LN(\boldsymbol\mu,\Sigma).
\]

The joint density of a multivariate log-normal distribution is
\[
f_{\mathbf x}(\mathbf x)=
\frac{1}{(2\pi)^{p/2}\,|\Sigma|^{1/2}}
\exp\!\Bigl[-\tfrac12(\ln\mathbf x-\boldsymbol\mu)^{\!\top}
            \Sigma^{-1}(\ln\mathbf x-\boldsymbol\mu)\Bigr]
\prod_{i=1}^{p}\frac1{x_i}.
\]

The marginal density of dimension $i$ is
\[
f_i(x_i)
=\frac1{x_i\,\sigma_i\sqrt{2\pi}}
  \exp\!\Bigl[-\tfrac12\Bigl(\tfrac{\ln x_i-\mu_i}{\sigma_i}\Bigr)^{\!2}\Bigr],
\qquad
Z_i \equiv \frac{\ln x_i-\mu_i}{\sigma_i}.
\]

\vspace{1em}
\paragraph{Step 2—Covariance matrix decomposition}

\[
\Sigma = D R D,
\qquad
D=\operatorname{diag}(\sigma_1,\dots,\sigma_p),
\qquad
R\text{ is the correlation matrix}.
\]
Then
\[
\Sigma^{-1}=D^{-1}R^{-1}D^{-1},
\qquad
|\Sigma| = \Bigl(\prod_{i=1}^{p}\sigma_i^2\Bigr)\,|R|.
\]

With the standardized variable
\(
Z_i \equiv \dfrac{\ln x_i-\mu_i}{\sigma_i},
\;
\mathbf Z=D^{-1}(\ln\mathbf x-\boldsymbol\mu),
\)
we have
\(
\operatorname{Cov}(\mathbf Z)
      = D^{-1}\Sigma D^{-1}=R.
\)
Hence the matrix \(R\) here corresponds exactly to the core parameter of the Gaussian copula.

\vspace{1.2em}
\paragraph{Step 3—Factorizing the joint density into \emph{marginals × copula}}
Because
\[
(\ln\mathbf x-\boldsymbol\mu)^{\!\top}\!\Sigma^{-1}(\ln\mathbf x-\boldsymbol\mu)
= \mathbf Z^{\!\top}R^{-1}\mathbf Z,
\]
we proceed as follows:

\begin{enumerate}
  \item Write the joint density from Step 1:  
    \[
      f_{\mathbf x}(\mathbf x)
      =\frac{1}{(2\pi)^{p/2}\,|\Sigma|^{1/2}}
        \exp\!\Bigl[-\tfrac12\,\mathbf Z^{\!\top}R^{-1}\mathbf Z\Bigr]
        \prod_{i=1}^{p}\frac1{x_i}.
    \]

  \item Use \(|\Sigma|^{1/2}=(\prod_i\sigma_i)|R|^{1/2}\) to separate the constant term, and decompose  
        \(\mathbf Z^{\!\top}R^{-1}\mathbf Z\) into
        \(\sum_i Z_i^2 + \mathbf Z^{\!\top}(R^{-1}-I)\mathbf Z\).

  \item For each dimension, add and subtract
        \(\tfrac12 Z_i^2\) to obtain the marginal log-normal density  
        \[
          f_i(x_i)=\frac{1}{x_i\,\sigma_i\sqrt{2\pi}}\,
                    \exp\!\bigl(-\tfrac12 Z_i^2\bigr).
        \]

  \item Re-assemble the terms:  
\[
\begin{aligned}
f_{\mathbf x}(\mathbf x)
&=
\underbrace{|R|^{-1/2}
            \exp\!\Bigl[-\tfrac12\,\mathbf Z^{\!\top}(R^{-1}-I)\mathbf Z\Bigr]}_{c_R(\mathbf u)}
\;\;
\underbrace{\prod_{i=1}^{p}
  \frac1{x_i\,\sigma_i\sqrt{2\pi}}
  \exp\!\bigl(-\tfrac12 Z_i^2\bigr)}_{\displaystyle\prod_i f_i(x_i)}.
\end{aligned}
\]
\end{enumerate}

Here
\(
u_i = F_i(x_i)=\Phi(Z_i)
\)
and
\[
c_R(\mathbf u)=
|R|^{-1/2}\,
\exp\!\Bigl[-\tfrac12\,\mathbf Z^{\!\top}(R^{-1}-I)\mathbf Z\Bigr]
\]
is exactly the density of the Gaussian copula.

\vspace{1em}
\paragraph{Step 4—Identify the Gaussian Copula density}

\[
c_R(\mathbf u)=
|R|^{-1/2}
\exp\!\Bigl[-\tfrac12\,\mathbf Z^{\!\top}(R^{-1}-I)\mathbf Z\Bigr],
\qquad
u_i = F_i(x_i)=\Phi(Z_i).
\]

Thus we arrive at the target formula
\[
\tilde b_{(k_1,k_2)}(\mathbf x)
  = |R_{(k_1,k_2)}|^{-1/2}
    \exp\!\Bigl[-\tfrac12\,\mathbf Z^{\!\top}
        \bigl(R_{(k_1,k_2)}^{\!-1}-I\bigr)\mathbf Z\Bigr]
    \prod_{i=1}^{p}\mathcal{LN}\!\bigl(x_i; \mu_{i,(k_1,k_2)},\sigma_{i,(k_1,k_2)}^{2}\bigr),
\]
showing that the two sides are identical.

\paragraph{Final conclusion}

\[
\boxed{
\widetilde b_{(k_1,k_2)}(\mathbf x)
  = |R_{(k_1,k_2)}|^{-1/2}
    \exp\!\Bigl[-\tfrac12\,\mathbf Z^{\!\top}
            \bigl(R_{(k_1,k_2)}^{-1}-I\bigr)\mathbf Z\Bigr]
    \prod_{i=1}^{p}\mathcal{LN}
           \!\bigl(x_{i}\!;\mu_{i,(k_1,k_2)},\sigma_{i,(k_1,k_2)}^{2}\bigr)
}
\]
Therefore, when the true distribution is multivariate log-normal, the combination of a Gaussian copula with log-normal marginals is lossless and unbiased.
% ------------------------------------------------------------
\subsection{Upper Bound on the Approximation Error of the Normalization Constant for the Zero-Inflated Log-Normal \emph{Emission} Distribution}
\label{subsec:ZI-LN-bound-revised}

\paragraph{1.\quad Notation and Model Specification}

\begin{itemize}
    \item \textbf{Zero-inflated density}\;
    \(f_{Z}(x_{\text{vis}})=\pi_{0}\,\mathbf 1_{x_{\text{vis}}=10}
    +(1-\pi_{0})\,f_{\mathrm{LN}}
    (x_{\text{vis}};\mu_{\text{vis}},\sigma_{\text{vis}}^{2}),\;
    0<\pi_{0}<1\).
    Here
    \(f_{\mathrm{LN}}(x_{\text{vis}};\mu_{\text{vis}},\sigma_{\text{vis}}^{2})
    =\dfrac{1}{x_{\text{vis}}\,\sigma_{\text{vis}}\sqrt{2\pi}}
    \exp\!\bigl\{-\tfrac{(\ln x_{\text{vis}}-\mu_{\text{vis}})^{2}}
    {2\sigma_{\text{vis}}^{2}}\bigr\},\;0<x_{\text{vis}}\leq10\).
    \item \textbf{Weighted normalization constant}\;
    \(C_{\mathrm{exact}}(w)=\int_{0}^{\infty}[f_{Z}(x_{\text{vis}})]^{w}\,dx,\;w>0\).
\end{itemize}

\paragraph{2.\quad Challenges and Choice of Approximation Strategy}

For an integral of the form \(C_{\mathrm{exact}}(w)=\int [a(x_{\text{vis}})+b(x_{\text{vis}})]^w dx\), an exact analytical solution is usually infeasible. 
Over-simplified approximations such as \((a+b)^w \approx a^w+b^w\) introduce non-negligible errors under our parameter settings.

We therefore adopt a higher-accuracy \emph{hybrid approximation} strategy, decomposing the integral into the sum of two main contributions:
\begin{enumerate}
    \item \textbf{Interaction term \(C_{\mathrm{interact}}(w)\)}: the contribution arising from the dominant point mass \(\pi_0\) and its interaction with the continuous density \(f_{\mathrm{LN}}(10)\) at \(x_{\text{vis}}=10\).
    \item \textbf{Body term \(C_{\mathrm{body}}(w)\)}: the contribution from the continuous log-normal part over the whole domain.
\end{enumerate}
The resulting approximation is \(C_{\mathrm{exact}}(w) \approx C_{\mathrm{hybrid}}(w) = C_{\mathrm{interact}}(w) + C_{\mathrm{body}}(w)\).

\begin{lstlisting}[language=Python]
# --- High-precision hybrid approximation of C(w) ---
d = lognorm(s=sigma, scale=np.exp(mu)).pdf(max_vis)     # f_LN(c)
T1 = w * (pi0 ** (w - 1)) * (1 - pi0) * d
T2 = 0.5 * w * (w - 1) * (pi0 ** (w - 2)) * ((1 - pi0) ** 2) * (d ** 2)
C_interact = (pi0 ** w) + T1 + T2                      # interaction term

ln_C_LN_PDF = log_norm_feat(w, mu, sigma)              # ln integral f_LN(x)^w dx
ln_C_body = w * np.log(1 - pi0) + ln_C_LN_PDF
C_body = np.exp(ln_C_body)                             # body term

C_hybrid = C_interact + C_body
logC[e] = np.log(max(C_hybrid, EPS))
\end{lstlisting}

\paragraph{3.\quad Derivation of the High-Precision Hybrid Approximation}

\subparagraph{3.1\quad Derivation of the Interaction Term \(C_{\mathrm{interact}}(w)\)}
The interaction term captures the local behaviour near \(x_{\text{vis}}=10\).  
Let \(a=\pi_0\mathbf{1}_{x_{\text{vis}}=10}\) and \(b=(1-\pi_0)f_{\mathrm{LN}}(x)\).  
Applying the generalized binomial theorem \((a+b)^w \approx a^w + w a^{w-1}b + \frac{w(w-1)}{2}a^{w-2}b^2 + \dots\) and integrating the first three terms yields
\[
Z_{\mathrm{interact}}(w) \approx \pi_0^w + T_1 + T_2,
\]
where \(d = f_{\mathrm{LN}}(c; \mu_{\mathrm{vis}}, \sigma_{\mathrm{vis}}^2)\).  
The first- and second-order cross terms are
\begin{align}
T_1 &= w\,\pi_0^{w-1}(1-\pi_0)\,d, \\
T_2 &= \frac{w(w-1)}{2}\,\pi_0^{w-2}(1-\pi_0)^2\,d^2.
\end{align}
These two terms adequately describe the local contribution at \(x_{\text{vis}}=10\).

\subparagraph{3.2\quad Derivation of the Body Term \(Z_{\mathrm{body}}(w)\)}

The body term captures the contribution of the log-normal distribution itself:
\[
Z_{\mathrm{body}}(w) = (1-\pi_0)^w \, Z_{\mathrm{LN,PDF}}(w),
\]
where \(Z_{\mathrm{LN,PDF}}(w) = \int_0^\infty [ f_{\mathrm{LN}}(x; \mu_{\mathrm{vis}}, \sigma_{\mathrm{vis}}^2) ]^w \,dx_{\text{vis}}\).  
This integral has a closed-form expression:
\begin{equation} \label{eq:ZLN_PDF_integral_corrected}
\boxed{
\begin{aligned}
\ln Z_{\mathrm{LN,PDF}}(w) ={}& (1-w)\ln\sigma_{\mathrm{vis}} + \frac{1-w}{2}\ln(2\pi) - \frac{1}{2}\ln w \\
& - (w-1)\mu_{\mathrm{vis}} + \frac{(w-1)^2}{2w}\sigma_{\mathrm{vis}}^2.
\end{aligned}
}
\end{equation}
Note that the coefficient in front of the location parameter \(\mu_{\text{vis}}\) is negative because the change of variables \(x_{\text{vis}}=e^y\) introduces the Jacobian term \(e^y\), which combines with \(x_{\text{vis}}^w\) in the denominator, producing a linear term $-(w-1)y$ in the exponent.

\subparagraph{3.3\quad Final Hybrid Approximation}
Combining the two parts gives
\begin{equation} \label{eq:Z_hybrid_corrected}
\begin{aligned}
Z_{\mathrm{hybrid}}(w) ={}& \left( \pi_0^w + w\pi_0^{w-1}(1-\pi_0)d + \frac{w(w-1)}{2}\pi_0^{w-2}(1-\pi_0)^2 d^2 \right) \\
& + (1-\pi_0)^w \, \exp\bigl( \ln Z_{\mathrm{LN,PDF}}(w) \bigr).
\end{aligned}
\end{equation}

\paragraph{4.\quad Numerical Error Upper Bound of the New Approximation}

Using the parameter settings in Table~\ref{tab:typical-fixed}, the worst-case relative error between \(C_{\mathrm{hybrid}}\) and \(C_{\mathrm{exact}}\) is about \(1.3 \times 10^{-4}\).  
To be conservative, we set
\begin{equation}
\label{eq:relerr-conservative-result-final}
\boxed{\mathrm{rel.err}(w) \lesssim 2 \times 10^{-4}}
\end{equation}
This is over two orders of magnitude smaller than the 4-5 \% error of the naive approximation \((a+b)^w \approx a^w+b^w\).

\begin{table}[h]
\centering
\caption{Parameter settings used in this paper (fixed \(\pi_{0}=0.99\))}
\label{tab:typical-fixed}
\begin{tabular}{ccc}
\hline
Symbol & Value / range & Basis \\
\hline
\(\pi_{0}\) & 0.99 & Proportion of Clear weather with \(x_{\text{vis}}=10\) \\
\(1-\pi_{0}\) & 0.01 & Derived from \(\pi_{0}\) \\
\(\mu_{\text{vis}}\) & \(\ln 8\)-\(\ln 10\) & Historical visibility peak 8-10 km \\
\(\sigma_{\text{vis}}\) & 0.4-0.6 & Historical visibility std. dev. \\
\((x_{\text{vis}})_{\max}\) & 10 km & Visibility measurement upper limit \\
\(w\) & 0.19-0.35 & Range of mutual-information weights \\
\hline
\end{tabular}
\end{table}

\paragraph{5.\quad Impact of the Error on Inference and Optimisation}

\begin{itemize}
    \item \textbf{Cumulative log-likelihood bias}:  
    For a sequence of length \(T\approx10^3\), the total bias is \(T\varepsilon < 0.2\), negligible relative to the average log-likelihood gain (\(\approx5\)) per EM iteration.
    \item \textbf{Change in posterior probability \(\gamma_t(k)\)}:  
    The scaling factor \(\kappa_t\) lies in \([1-2\times10^{-4},\,1+2\times10^{-4}]\), giving \(|\Delta\gamma_t/\gamma_t| < 0.02\%\), three orders of magnitude smaller than the natural variability (20-40 \%).
    \item \textbf{Effect on gradient terms}:  
    Absolute gradient bias \(\le 2 \times 10^{-4}\).  
    Relative error \(\le 0.2\%\), well below typical optimiser thresholds (\(10^{-3}\)).
\end{itemize}

\begin{table}[h]
\centering
\caption{Time standard deviation of \(\gamma_{ij}(t)\) (percentage)}
\label{tab:gamma-std-inline}
\begin{tabular}{ccc}
\hline
Combination \((Z_1,Z_2)\) & Event interpretation & Std\(_t\)[\(\gamma_{ij}\)] \\
\hline
(0,0) & Clear & 38.9 \\
(0,1) & Dust  & 23.0 \\
(1,0) & Haze  & 33.5 \\
\hline
\end{tabular}
\end{table}

\paragraph{6.\quad Conclusion}

The simple approximation incurs large errors, but the high-precision hybrid approximation (\ref{eq:Z_hybrid_corrected}) is physically and mathematically sound.  
Under the settings in Table~\ref{tab:typical-fixed}, its relative error does not exceed \(2 \times 10^{-4}\), several orders of magnitude below the numerical noise in optimisation, and thus has no impact on the reliability or accuracy of inference.
% ------------------------------------------------------------
\subsection{Normalization of Emission Probabilities under a Global Weight}\label{subsec:global-temp-norm}

\paragraph{1.\quad Motivation and Theoretical Definition}

After applying the per-dimension weights \(w_e\), we obtain the weighted emission
probability \(b_{i,(k_1,k_2)}^{\langle w\rangle}(x_{i,t})\) defined in~\eqref{eq:weight}.  
To further control the sharpness of the joint
emission density, we introduce a positive global hyper-parameter \(v>0\).

\paragraph{Globally weighted emission probability}
For hidden state \((k_1,k_2)\) and observation vector \(\boldsymbol x_t\) we set  

\begin{equation}
    b_{(k_1,k_2)}^{\langle w,v\rangle}(\boldsymbol x_t) = \pr_{\text{Global Wtd}}(\boldsymbol x_t | \boldsymbol Z_t = \boldsymbol Z)=   \frac{\bigl[\Pr^{\text{Weighted}}(\boldsymbol x_t \mid \boldsymbol Z_t=(k_1,k_2))\bigr]^{1/v}}
       {C^{\text{GlobalWtd}}_{(k_1,k_2)}}
    \label{eq:temperature}
\end{equation}
where the normalising constant  

\begin{equation}
    C^{\text{Global Wtd}}_{(k_1,k_2)}(v)=\idotsint_{\mathbb{R}^{E}} \left[  \pr^{\text{Weighted}}(\boldsymbol {x}_t | \boldsymbol Z=(k_1,k_2)) \right] ^{1/v}d\boldsymbol x
\end{equation} 
ensures that \(b_{(k_1,k_2)}^{\langle w,v\rangle}(\cdot)\) integrates to one.

When \(v=1\) the formula reduces to the locally weighted density; larger \(v\) values
produce a flatter (“cooler”) distribution, whereas \(v<1\) sharpens it.

Since \(C^{\text{Global Wtd}}_{(k_1,k_2)}(v)\) lacks a closed form and is costly to compute, we seek a practical alternative.

\paragraph{2.\quad Practical Scheme: Summation over \(\boldsymbol Z\) instead of Integration over \(\boldsymbol x\)}

In conditional inference (Viterbi, EM, etc.), we require \(\Pr(\boldsymbol Z_t=\boldsymbol Z \mid \boldsymbol {x}_t)\) for a fixed observation \(\boldsymbol {x}_t\).
\begin{itemize}
    \item \textbf{Key idea}: For a given \(t\) and \(\boldsymbol {X}_t\), the posterior probabilities over all \(Z\in\mathcal{Z}\) must sum to 1. We can exploit this to normalise without computing the integral.
    \item \textbf{Steps}:  
    Compute the unnormalised scores \(N_Z := \widehat{P}_{v}(\boldsymbol {x}_t\mid \boldsymbol Z)\) for all \(Z\).  
    The normalising factor is
    \[
      C^{\text{Global Wtd}}_{(k_1,k_2)}(v) = \sum_{\boldsymbol Z' \in \mathcal{\boldsymbol Z}} N_{\boldsymbol Z'}
    \]
    and the exact posterior is
    \[
      \Pr(\boldsymbol Z_t=\boldsymbol Z \mid \boldsymbol {x}_t) = \frac{N_{\boldsymbol Z}}{C^{\text{Global Wtd}}_{(k_1,k_2)}(v)}.
    \]
    \item \textbf{Equivalence}: In Viterbi-style max-product computations, the common constant \(C_{\text{global}}\) cancels, so the decision is unchanged. Hence this summation-based normalisation is algorithmically equivalent to the integral-based definition.
\end{itemize}

\paragraph{3.\quad Numerical Accuracy}

\begin{itemize}
    \item \textbf{Single-dimension}: The visibility dimension uses an approximation with relative error \(<4\times10^{-4}\) (see~\eqref{eq:Z_hybrid_corrected}); other dimensions are exact.
    \item \textbf{Global}: Summation over the finite set \(\mathcal{Z}\) is algebraically exact.
    \item \textbf{Overall}: The dominant error remains \(O(10^{-4})\), negligible for log-likelihoods, posteriors, and gradients.
\end{itemize}

% \paragraph{4.\quad Summary}

% The practical global-weight normalisation proceeds by \emph{(i) per-dimension weighting and multiplication, (ii) power transformation, and (iii) normalisation over the hidden-state vector}.  
% This replaces an intractable integral with a simple vector summation while remaining mathematically equivalent for inference and optimisation, achieving an ideal trade-off between numerical precision and computational efficiency
% ------------------------------------------------------------
\subsection{Comparison of Classification Performance across Models}
\label{sec:cm_compare_en}
%------------------------------------------------------------

To provide an intuitive assessment of how the candidate models differ on the three-class task “Clear / Haze / Dust”, this section presents the confusion matrices in the order of model evolution and gives a quantitative interpretation.

\paragraph{Overview of Model Configurations}

Table~\ref{tab:model_configurations_en} summarises, for each model, the differences in \emph{correlation structure}, \emph{observation weight}, and \emph{global optimisation}, outlining the technical progression from the baseline to the final version.

\begin{table}[H]
  \centering
  \renewcommand{\arraystretch}{1.4} % 调整行高
  \caption{Detailed comparison of model configurations}
  \label{tab:model_configurations_en}
  \begin{tabularx}{\textwidth}{@{}l l l c X@{}}
    \toprule
    \textbf{Code} & \textbf{Correlation structure} & \textbf{Obs.~weight $w$} & \textbf{Global opt.} & \textbf{Key description} \\
    \midrule
    M0  & None (independence) & None            & No  & Baseline FHMM; ignores inter-dimensional correlation. \\
    \addlinespace
    M1a & Joint log-normal                & None            & No  & Uses a covariance matrix to capture \emph{linear} correlation. \\
    M1b & Joint log-normal                & Normalised MI & No  & Adds mutual-information weights on top of M1a. \\
    \addlinespace
    M2a & Gaussian copula                 & None            & No  & Uses a copula to capture \emph{non-linear} correlation. \\
    M2b & Gaussian copula                 & Raw MI        & No  & Adds unnormalised MI weights on top of M2a. \\
    M2c & Gaussian copula                 & Normalised MI & No  & Adds normalised MI weights on top of M2a. \\
    \addlinespace
    M2d & Gaussian copula & Normalised MI & Yes & Final optimised model: jointly tunes $\Sigma w$ and the global weight $v$. \\
    \bottomrule
  \end{tabularx}
\end{table}

\paragraph{Interpretation of Performance Evolution}

\begin{figure}[H]
  \centering
  \includegraphics[width=0.9\textwidth]{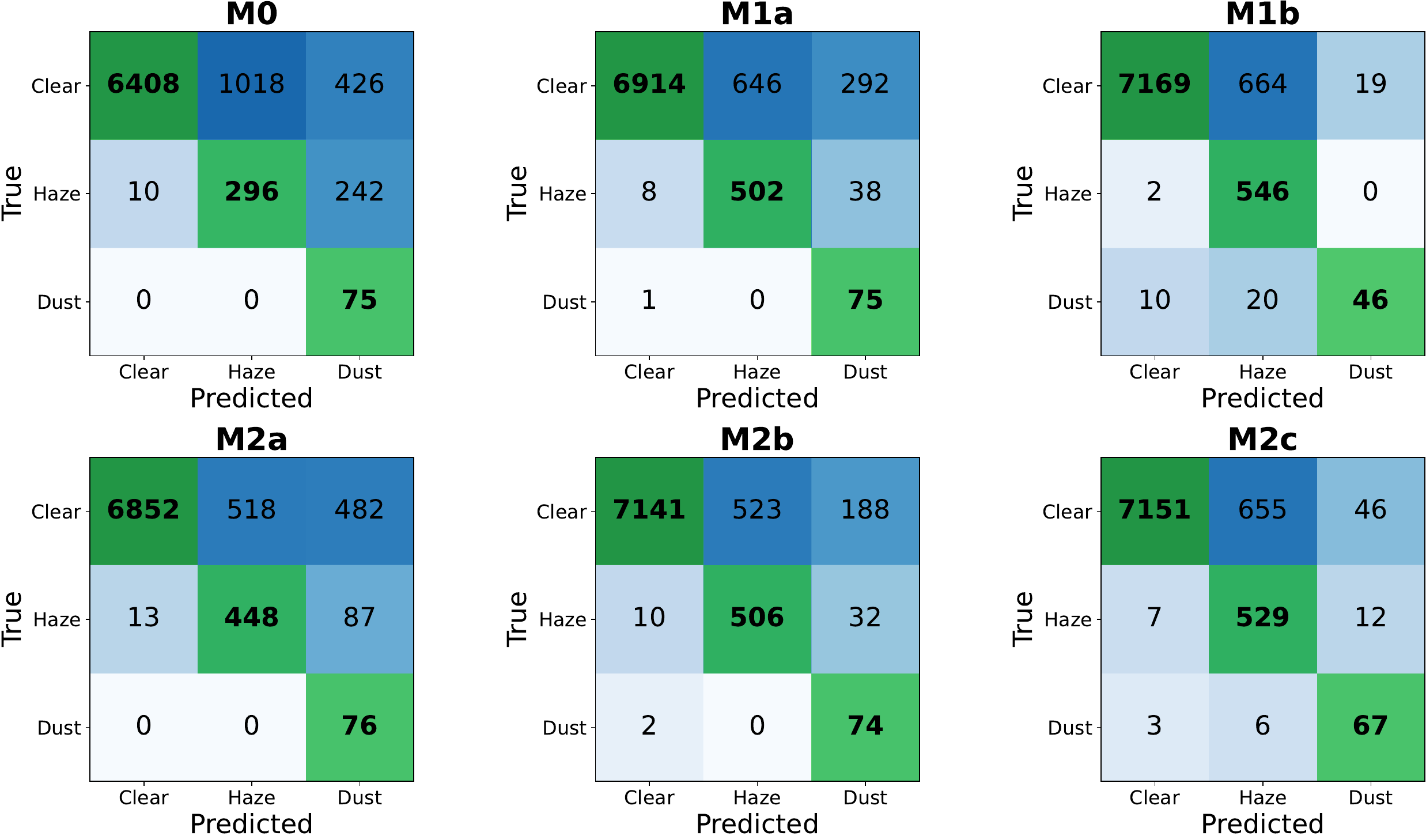}
  \caption{Confusion-matrix comparison across different models}
  \label{fig:cm_overall}
\end{figure}

\begin{figure}[H]
  \centering
  \includegraphics[width=0.28\textwidth]{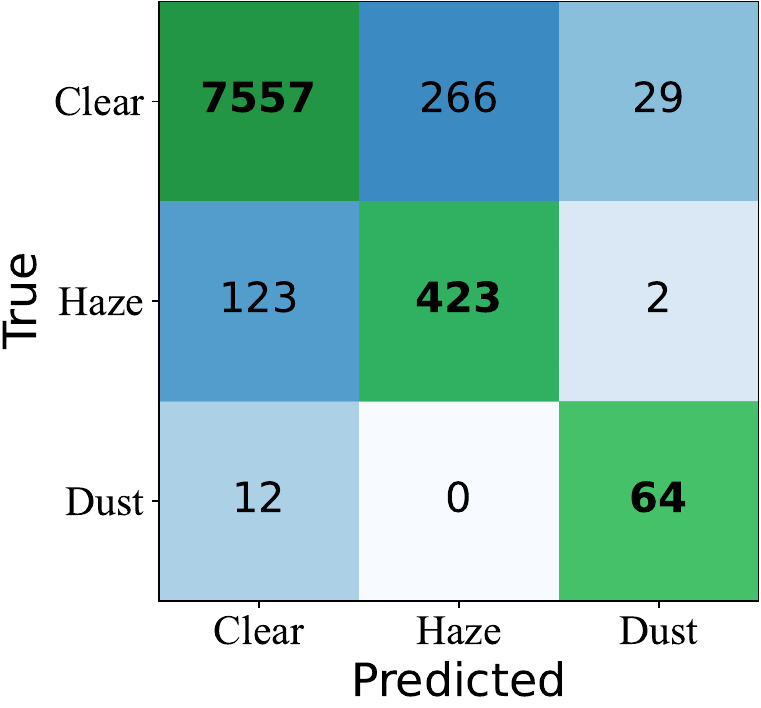}
  \caption{Confusion matrix of the final, parameter-optimised model}
  \label{fig:final_confusion_matrix}
\end{figure}

Figures~\ref{fig:cm_overall} and~\ref{fig:final_confusion_matrix} compile the confusion matrices of all the models discussed above and show that performance improves steadily as correlation structures and weighting mechanisms are progressively introduced.

\paragraph{Evaluation of the Final Optimised Model}

Starting from \texttt{M2c}, a joint grid search over the global weight \( v \) and the sum of observation weights \( \Sigma w \) yields the final model \texttt{M2d}. Its confusion matrix is shown in Figure~\ref{fig:final_confusion_matrix}.

Compared with the other models in Figure~\ref{fig:cm_overall}, \texttt{M2d} attains the highest true-positive rates on all three diagonals and reduces off-diagonal errors to the minimum. The improvement is especially notable for the rare class “Dust” (see Table~\ref{tab:model_comparison_professional}; F1 increases to 0.7077). These results demonstrate that \emph{jointly tuning the scale of observation weights and the global weight within the copula framework effectively balances performance between majority and rare classes.}

% ------------------------------------------------------------
\subsection{Final Parameters of All Models}

\subsubsection*{Final Parameters of Model M0}

%=================== Transition matrices (updated data) ===================
\begin{table}[htbp]
  \centering
  \small
  \caption{Final transition matrices of model M0}

  \subfloat[Haze]{%
    \begin{tabular}{c|cc}
      \toprule
      \diagbox{from}{to} & 0 & 1 \\
      \midrule
      0 & 0.9833 & 0.0886 \\
      1 & 0.0167 & 0.9114 \\
      \bottomrule
    \end{tabular}}
  \qquad
  \subfloat[Dust]{%
    \begin{tabular}{c|cc}
      \toprule
      \diagbox{from}{to} & 0 & 1 \\
      \midrule
      0 & 0.9926 & 0.0765 \\
      1 & 0.0074 & 0.9235 \\
      \bottomrule
    \end{tabular}}
\end{table}

%=================== μ and σ (updated data, two decimals) ===================
\begin{table}[H]
    \centering
    \caption{Final means \( \mu_{hk} \) and standard deviations \( \sigma_{hk} \) for model~M0}
    \label{tab:mu_sigma_M0}

    \sisetup{table-number-alignment = center}
    \begin{tabular}{
        @{} c c
        S[table-format=1.2] % PM10
        S[table-format=1.2] % Wind
        S[table-format=1.2] % Visibility
        S[table-format=1.2] % Humidity
        @{}
    }
        \toprule
        \multirow{2}{*}{Hidden state $(k_1,k_2)$} &
        \multirow{2}{*}{Statistic} &
        \multicolumn{4}{c}{Observed variables} \\
        \cmidrule(l){3-6}
         & & {PM$_{10}$} & {Wind} & {Visibility} & {Humidity} \\
        \midrule

        \multirow{2}{*}{$(0,0)$}
        & $\mu$    & 3.68 & 0.93 & 2.04 & 3.72 \\
        & $\sigma$ & 0.76 & 0.67 & 0.10 & 0.57 \\
        \addlinespace

        \multirow{2}{*}{$(0,1)$}
        & $\mu$    & 5.10 & 0.81 & 1.69 & 3.90 \\
        & $\sigma$ & 0.93 & 0.67 & 0.41 & 0.65 \\
        \addlinespace

        \multirow{2}{*}{$(1,0)$}
        & $\mu$    & 4.54 & 0.33 & 1.96 & 4.44 \\
        & $\sigma$ & 0.38 & 0.41 & 0.43 & 0.12 \\

        \bottomrule
    \end{tabular}
\end{table}

M0 does not model correlations among the observed variables; therefore, no correlation matrix is provided.

\subsubsection*{Final Parameters Shared by Models M1a and M1b}

%=================== Transition matrices (updated data) ===================
\begin{table}[htbp]
  \centering
  \small
  \caption{Final transition matrices of models M1a and M1b}
  \label{tab:fhmm_transitions_M1ab}

  \subfloat[Haze]{%
    \begin{tabular}{c|cc}
      \toprule
      \diagbox{from}{to} & 0 & 1 \\
      \midrule
      0 & 0.9873 & 0.0784 \\
      1 & 0.0127 & 0.9216 \\
      \bottomrule
    \end{tabular}}
  \qquad
  \subfloat[Dust]{%
    \begin{tabular}{c|cc}
      \toprule
      \diagbox{from}{to} & 0 & 1 \\
      \midrule
      0 & 0.9964 & 0.0722 \\
      1 & 0.0036 & 0.9278 \\
      \bottomrule
    \end{tabular}}
\end{table}

%=================== μ and σ (updated data, two decimals) ===================
\begin{table}[H]
    \centering
    \caption{Final means \( \mu_{hk} \) and standard deviations \( \sigma_{hk} \) for models~M1a and M1b}
    \label{tab:mu_sigma_M1ab}

    \sisetup{table-number-alignment = center}
    \begin{tabular}{
        @{} c c
        S[table-format=1.2] % PM10
        S[table-format=1.2] % Wind
        S[table-format=1.2] % Visibility
        S[table-format=1.2] % Humidity
        @{}
    }
        \toprule
        \multirow{2}{*}{Hidden state $(k_1,k_2)$} &
        \multirow{2}{*}{Statistic} &
        \multicolumn{4}{c}{Observed variables} \\
        \cmidrule(l){3-6}
         & & {PM$_{10}$} & {Wind} & {Visibility} & {Humidity} \\
        \midrule

        \multirow{2}{*}{$(0,0)$}
        & $\mu$    & 3.74 & 0.87 & 2.08 & 3.78 \\
        & $\sigma$ & 0.76 & 0.67 & 0.01 & 0.56 \\
        \addlinespace

        \multirow{2}{*}{$(0,1)$}
        & $\mu$    & 5.55 & 1.10 & 1.87 & 3.33 \\
        & $\sigma$ & 0.61 & 0.74 & 0.39 & 0.73 \\
        \addlinespace

        \multirow{2}{*}{$(1,0)$}
        & $\mu$    & 4.58 & 0.47 & 1.68 & 4.41 \\
        & $\sigma$ & 0.65 & 0.48 & 0.43 & 0.17 \\

        \bottomrule
    \end{tabular}
\end{table}

Unlike the Gaussian-copula models, the joint log-normal models cannot accommodate the inflated mixture distribution of Visibility under Clear conditions. Consequently, an independence assumption is adopted; that is, \(R_{(0,0)}\) is the identity matrix. For the remaining hidden states, the global correlation matrix is still independent of the hidden state and depends only on the types of observed variables.

\begin{figure}[H]
\centering
\includegraphics[width=0.6\textwidth]{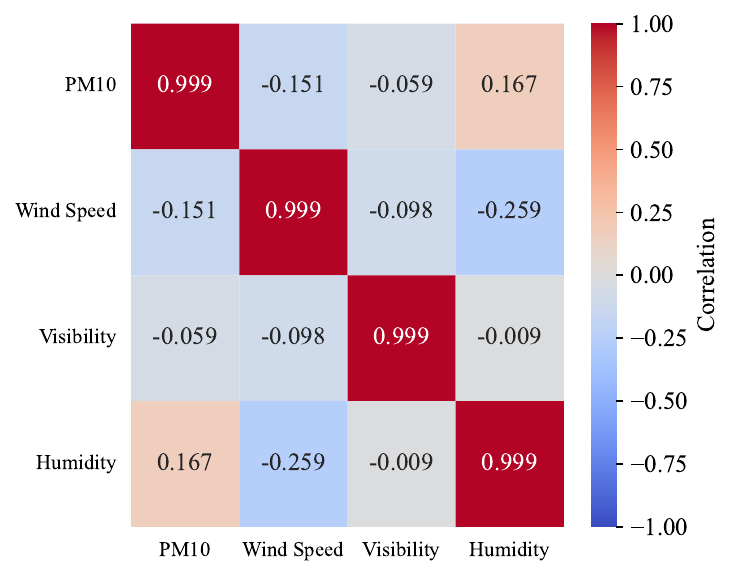}
\caption{Heat map of the final global correlation matrix for models M1a and M1b}
\label{fig:corr_heatmap_M1}
\end{figure}

\subsubsection*{Final Parameters Shared by Models M2a, M2b, M2c, and M2d}

%===== Transition matrices =====
\begin{table}[htbp]
  \centering
  \small
  \caption{Final transition matrices shared by models M2a, M2b, M2c, and M2d (LNC family)}
  \label{tab:fhmm_transitions_M2abcd}

  \subfloat[Haze]{%
    \begin{tabular}{c|cc}
      \toprule
      \diagbox{from}{to} & 0 & 1 \\
      \midrule
      0 & 0.9887 & 0.0830 \\
      1 & 0.0113 & 0.9170 \\
      \bottomrule
    \end{tabular}}
  \qquad
  \subfloat[Dust]{%
    \begin{tabular}{c|cc}
      \toprule
      \diagbox{from}{to} & 0 & 1 \\
      \midrule
      0 & 0.9937 & 0.0762 \\
      1 & 0.0063 & 0.9238 \\
      \bottomrule
    \end{tabular}}
\end{table}

\begin{table}[H]
    \centering
    \caption{Final means \( \mu_{hk} \) and standard deviations \( \sigma_{hk} \) shared by models M2a, M2b, M2c, and M2d}
    \label{tab:mu_sigma_M2abcd}
    
    \sisetup{table-number-alignment = center}
    \begin{tabular}{
        @{}
        c
        c
        S[table-format=1.4]
        S[table-format=1.4]
        S[table-format=1.4]
        S[table-format=1.4]
        @{}
    }
        \toprule
        \multirow{2}{*}{Hidden state $(k_1,k_2)$} & 
        \multirow{2}{*}{Statistic} & 
        \multicolumn{4}{c}{Observed variables} \\
        \cmidrule(l){3-6}
         & & {PM$_{10}$} & {Wind} & {Visibility} & {Humidity} \\
        \midrule
        
        \multirow{2}{*}{$(0,0)$} 
        & $\mu$    & 3.73 & 0.87 & 2.04 & 3.79 \\
        & $\sigma$ & 0.76 & 0.67 & 0.10 & 0.56 \\ 
        \addlinespace
        
        \multirow{2}{*}{$(0,1)$} 
        & $\mu$    & 5.33 & 0.96 & 1.90 & 3.53 \\
        & $\sigma$ & 0.60 & 0.72 & 0.36 & 0.72 \\
        \addlinespace
        
        \multirow{2}{*}{$(1,0)$} 
        & $\mu$    & 4.50 & 0.49 & 1.65 & 4.45 \\
        & $\sigma$ & 0.67 & 0.49 & 0.44 & 0.13 \\
        
        \bottomrule
    \end{tabular}
\end{table}

\begin{figure}[H]
\centering
\includegraphics[width=0.6\textwidth]{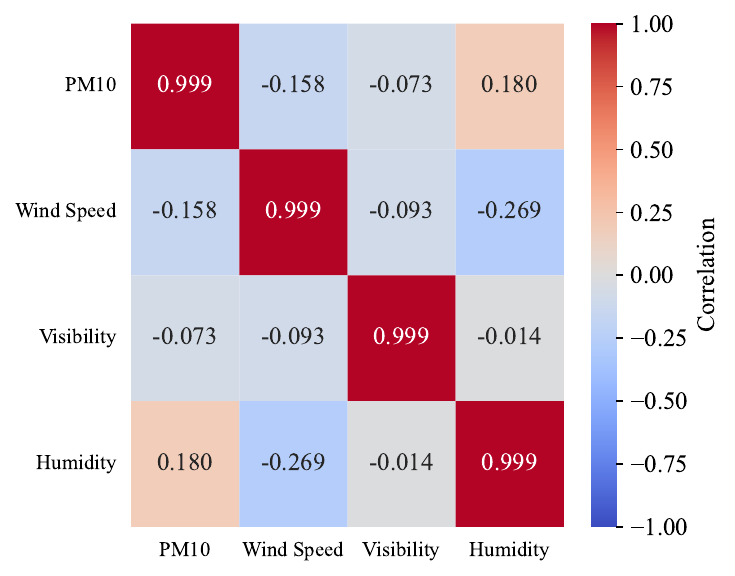}
\caption{Heat map of the final global correlation matrix for models M2a, M2b, M2c, and M2d}
\label{fig:corr_heatmap_M2d}
\end{figure}
% ------------------------------------------------------------
\subsection{Fast Algorithms for Two-Chain FHMM}
\label{subsec:fast-2chain-fhmm}

Let the hidden-state vector be
\[
  \boldsymbol Z_t=(Z_t^{(1)},Z_t^{(2)})\equiv (k_1,k_2),
  \qquad k_1,k_2\in\{1,\dots,K\}.
\]
Each chain has its own transition matrix
\(A^{(1)},A^{(2)}\in\mathbb R^{K\times K}\)
and initial distribution
\(\pi^{(1)},\pi^{(2)}\in\mathbb R^{K}\).
The joint transition matrix is the Kronecker product
\[
  A = A^{(1)} \otimes A^{(2)}, \qquad 
  A_{(k_1,k_2),(j_1,j_2)} = A^{(1)}_{k_1j_1}\,A^{(2)}_{k_2j_2}.
\]

% ----------------------------------------------------------------
\subsubsection*{Forward Probabilities}

Define the unscaled forward tensor
\[
  \alpha_t(k_1,k_2)=
  \Pr\!\bigl(x_{0:t},Z_t^{(1)}=k_1,Z_t^{(2)}=k_2\bigr).
\]
Scaling avoids underflow:
\(c_t=\sum_{k_1,k_2}\alpha_t(k_1,k_2)\),\;
\(\tilde\alpha_t=\alpha_t/c_t\).
Recursion
\[
  \alpha_t(k_1,k_2)=
  B_{k_1,k_2}(\bm x_t)
  \sum_{j_1,j_2}
  A^{(1)}_{k_1j_1}A^{(2)}_{k_2j_2}\,
  \tilde\alpha_{t-1}(j_1,j_2),
\]
which factors into two \(\mathcal O(K^3)\) contractions.

\begin{algorithm}[H]
\caption{\textsc{Forward\_2Chain} — tensorised forward pass}
\label{alg:fwd2}
\begin{algorithmic}[1]
  \Require Observations $\bm x_{0:T-1}$;
           $A^{(1)},A^{(2)}$; $\pi^{(1)},\pi^{(2)}$;
           \emph{two-step} emission tensor $B(\cdot)$
  \Ensure  $\{\tilde\alpha_t\}$ and log-likelihood $\log L$
  \State $\alpha_0\gets(\pi^{(1)}\!\otimes\!\pi^{(2)})\odot B(\bm x_0)$
  \State $c_0\gets\text{sum}(\alpha_0)$;\;
         $\tilde\alpha_0\gets\alpha_0/c_0$;\;
         $\log L\gets\log c_0$
  \For{$t=1$ \textbf{to} $T-1$}
     \State $\mathbf V\gets\tilde\alpha_{t-1}$
     \State $\mathbf V\gets\text{einsum}(A^{(1)},\mathbf V)$
     \State $\mathbf V\gets\text{einsum}(A^{(2)},\mathbf V)$
     \State $\alpha_t\gets B(\bm x_t)\odot\mathbf V$
     \State $c_t\gets\text{sum}(\alpha_t)$;\;
            $\tilde\alpha_t\gets\alpha_t/c_t$;\;
            $\log L\gets\log L+\log c_t$
  \EndFor
  \Return $\log L$, $\{\tilde\alpha_t\}$
\end{algorithmic}
\end{algorithm}

% ----------------------------------------------------------------
\subsubsection*{Backward Probabilities}

The backward probability is
\(
\beta_t(k_1,k_2)=
 \Pr(x_{t+1:T-1}\mid Z_t^{(1)}=k_1, Z_t^{(2)}=k_2)
\),
scaled in the same way with \(c_{t+1}\):
\(
\tilde\beta_t=\beta_t/c_{t+1}
\).
The recursion reads
\[
\tilde\beta_t(k_1,k_2)=
\frac{1}{c_{t+1}}
\sum_{j_1,j_2}
A^{(1)}_{j_1k_1}A^{(2)}_{j_2k_2}\,
B_{j_1,j_2}(x_{t+1})\,
\tilde\beta_{t+1}(j_1,j_2).
\]

\begin{algorithm}[H]
\caption{\textsc{Backward\_2Chain} — tensorised backward pass}
\label{alg:bwd2}
\begin{algorithmic}[1]
  \Require Same inputs as Alg.~\ref{alg:fwd2}; scale factors $\{c_t\}$
  \Ensure $\{\tilde\beta_t\}$
  \State $\tilde\beta_{T-1}\gets\mathbf 1_{K\times K}$
  \For{$t=T-2$ \textbf{downto} $0$}
     \State $\mathbf V\gets B(\bm x_{t+1})\odot\tilde\beta_{t+1}$
     \State $\mathbf V\gets\text{einsum}(A^{(1)\top},\mathbf V)$
     \State $\mathbf V\gets\text{einsum}(A^{(2)\top},\mathbf V)$
     \State $\tilde\beta_t\gets\mathbf V/\,c_{t+1}$
  \EndFor
  \Return $\{\tilde\beta_t\}$
\end{algorithmic}
\end{algorithm}

\textbf{Complexity.}  
Forward / backward each need two tensor-matrix contractions per step:
\(\mathcal O(TK^3)\) time, \(\mathcal O(K^2)\) memory.

% ----------------------------------------------------------------
\subsubsection*{Viterbi Decoding (with MI Weights \& Global Weight)}

Let the log-Viterbi score be  
\[
  \omega_t(k_1,k_2)=
  \log\max_{\bm Z_{0:t-1}}
  \Pr\!\bigl(\bm Z_{0:t},\bm x_{0:t}\bigr).
\]

\begin{enumerate}
  \item \textbf{Transition step}\;  
        To avoid forming a \(K^4\) matrix, perform two sequential maximisations:
        \[
          T_1(j_1,k_2)=
            \max_{i_1}\Bigl[
              \omega_{t-1}(i_1,k_2)+\log A^{(1)}_{j_1i_1}
            \Bigr],
        \qquad
          T_2(k_1,k_2)=
            \max_{i_2}\Bigl[
              T_1(k_1,i_2)+\log A^{(2)}_{k_2i_2}
            \Bigr].
        \]

\item \textbf{Two-step weighted emission}\;
  \begin{enumerate}[label=(\alph*)]
    \item \emph{Local MI weighting}\;  
          \[
            b^{\langle w\rangle}_{k_1,k_2}(\bm x_t)=
            \prod_{i=1}^{E}
              \frac{p_i(x_{i,t}\mid k_1,k_2)^{\,w_{i,(k_1,k_2)}}}
                   {C^{\text{Weighted}}_{i,(k_1,k_2)}} ,
            \qquad
            \log b^{\langle w\rangle}_{k_1,k_2} =
            \sum_{i=1}^{E}
              w_{i,(k_1,k_2)}\,
              \log p_i(x_{i,t}\mid k_1,k_2).
          \]
    \item \emph{Global weight \(v\) and normalisation}\;  
          \[
            S_{k_1,k_2}= \frac{1}{v}\,\log b^{\langle w\rangle}_{k_1,k_2},
            \quad
            \widetilde C^{\text{Global}}(v,\bm x_t)=
              \sum_{j_1,j_2}\exp\bigl(S_{j_1,j_2}\bigr),
          \]
          \[
            E_t(k_1,k_2)=
              \frac{1}{v}\,\log b^{\langle w\rangle}_{k_1,k_2}
              -\log\widetilde C^{\text{Global}}(v,\bm x_t),
          \]
          i.e.\ \(E_t\) is the log-softmax of
          \(\tfrac1v\log b^{\langle w\rangle}\) over all hidden states,
          which guarantees
          \(\sum_{k_1,k_2} \exp\bigl(E_t(k_1,k_2)\bigr)=1\).
  \end{enumerate}

  \item \textbf{Recursion}\;
        \[
          \omega_t(k_1,k_2)=
          T_2(k_1,k_2)+E_t(k_1,k_2).
        \]
\end{enumerate}

\begin{algorithm}[H]
\caption{\textsc{Viterbi\_2Chain\_Weighted}}
\label{alg:viterbi2}
\begin{algorithmic}[1]
  \Require $\bm x_{0:T-1}$; $\log A^{(1)},\log A^{(2)}$;
          $\pi^{(1)},\pi^{(2)}$; weights $W$; global weight $v$
  \Ensure  MAP path $\{\hat k_1(t),\hat k_2(t)\}$
  \State Compute $E_0$ using the two-step rule; 
         $\omega_0\gets\log(\pi^{(1)}\!\otimes\!\pi^{(2)})+E_0$
  \For{$t=1$ \textbf{to} $T-1$}
     \State Compute $T_1$ and $T_2$ (transition max)
     \State $\log b^{\langle w\rangle}\gets
            \sum_e W\,\log p_e(\cdot)$
     \State $S\gets \log b^{\langle w\rangle}/v$
     \State $E_t\gets S-\text{LogSumExp}(S)$
     \State $\omega_t\gets T_2+E_t$
  \EndFor
  \State Back-track via stored pointers to obtain the path
\end{algorithmic}
\end{algorithm}

\paragraph{Complexity.}
Each step still needs only two max-contractions; the extra
vectorised softmax is \(\mathcal O(K^2)\).
Overall:
\(\mathcal O(TK^3)\) time, \(\mathcal O(K^2)\) space.

% ============================================================

% ------------------------------------------------------------
\subsection*{Common Initial Parameters for All Models}

\begin{table}[H]
  \centering
  \small
  \caption{Initial transition-probability matrices estimated from historical data}
  \label{tab:fhmm_transitions}
  % -------- Haze ----------
  \subfloat[Haze]{%
    \begin{tabular}{c|cc}
      \toprule
      \diagbox{from}{to} & 0 & 1 \\ \midrule
      0 & 0.9884 & 0.0116 \\     % row 0
      1 & 0.1330 & 0.8670 \\     % row 1
      \bottomrule
    \end{tabular}}
  \qquad
  % -------- Dust ----------
  \subfloat[Dust]{%
    \begin{tabular}{c|cc}
      \toprule
      \diagbox{from}{to} & 0 & 1 \\ \midrule
      0 & 0.9998 & 0.00023 \\    % row 0
      1 & 0.1250 & 0.8750 \\     % row 1
      \bottomrule
    \end{tabular}}
\end{table}

\begin{table}[H]
  \centering
  \caption{Initial means \( \mu_{hk} \) and standard deviations \( \sigma_{hk} \) (obtained via K-means) for each hidden state}
  \sisetup{table-number-alignment = center}

  \begin{tabular}{@{} c c
                   S[table-format=1.4] S[table-format=1.4]
                   S[table-format=1.4] S[table-format=1.4] @{}}
    \toprule
    \multirow{2}{*}{Hidden state $(k_1,k_2)$} &
    \multirow{2}{*}{Statistic} &
    \multicolumn{4}{c}{Observed variables} \\ \cmidrule(l){3-6}
      & 
      & \multicolumn{1}{c}{PM$_{10}$}
      & \multicolumn{1}{c}{Wind}
      & \multicolumn{1}{c}{Visibility}
      & \multicolumn{1}{c}{Humidity} \\ \midrule

    % ---------- (0,0) ----------
    \multirow{2}{*}{$(0,0)$}
      & $\mu$    & 2.8276 & 0.3483 & 1.9637 & 3.6430 \\
      & $\sigma$ & 0.3794 & 0.3466 & 0.1163 & 0.2063 \\ \addlinespace
    % ---------- (0,1) ----------
    \multirow{2}{*}{$(0,1)$}
      & $\mu$    & 4.9151 & 1.4853 & 1.4630 & 2.8348 \\
      & $\sigma$ & 0.4478 & 0.3875 & 0.1400 & 0.3234 \\ \addlinespace
    % ---------- (1,0) ----------
    \multirow{2}{*}{$(1,0)$}
      & $\mu$    & 3.9571 & -4.6052 & 0.9672 & 4.3288 \\
      & $\sigma$ & 0.2908 & 0.0100 & 0.2055 & 0.1840 \\ \addlinespace
    % ---------- (1,1) ----------
    \multirow{2}{*}{$(1,1)$}
      & $\mu$    & 3.8999 & -0.9239 & 1.4646 & 3.6022 \\
      & $\sigma$ & 0.3782 & 0.3002 & 0.1585 & 0.2456 \\

    \bottomrule
  \end{tabular}
\end{table}

The correlation matrices are initialised to the identity matrix to avoid excessively strong parameter coupling at the outset, which could otherwise cause the EM algorithm to converge rapidly to poor local maxima.

% ------------------------------------------------------------
\subsection{Notation}
\begin{itemize}
    \item $\beta(\boldsymbol C_t)$: backward probability vector
    \item $\boldsymbol A$: state-transition matrix
    \item $\boldsymbol X_t$: observation at time~$t$
    \item $c_t$: normalisation constant
    \item $\omega_t$: Viterbi recursion score
    \item $\boldsymbol b_t$: observation-probability vector
    \item $\boldsymbol w_{i,(k_1,k_2)}$: feature-weight matrix for the observation chains
    \item $v$: global weight
    \item $M$: number of hidden chains
    \item $K$: number of states per hidden chain
    \item $E$: number of observed features
\end{itemize}

\section*{Funding}

This research is supported by National Natural Science Foundation of China (T2293773).

\section*{Data availability}

Hourly meteorological observations for Beijing Capital International Airport (station code: ZBAA) were downloaded from the RP5 weather service (https://rp5.ru/).  Hourly air-quality data for the nearby Beijing Olympic Sports Center site (station code: 1011A) were obtained from the China National Comprehensive Earth Observation Data Sharing Platform (https://www.chinageoss.cn/) and the National Urban Air-Quality Real-time Publishing Platform of the China National Environmental Monitoring Center (https://air.cnemc.cn:18007/). The processed dataset supporting the findings of this study is available from the corresponding author upon reasonable request.

\bibliographystyle{plainnat}
\bibliography{main}
\end{document}